\documentclass[]{aa}  
\usepackage{booktabs}
\usepackage{graphicx}
\usepackage{verbatim}
\usepackage{breqn}
\usepackage{txfonts}
\usepackage{subfigure}
\usepackage{gensymb}
\usepackage{natbib,twoopt}

\graphicspath{{/media/alberto/TRANSCEND/masini15/figures/},{/Volumes/TRANSCEND/masini15/figures/}}
\bibpunct{(}{)}{;}{a}{}{,} 

\begin{document} 
\title{\textit{NuSTAR} observations of water megamaser AGN}

\author{A. Masini\inst{1,2}, A. Comastri\inst{1}, M. Balokovi\'c\inst{3}, I. Zaw\inst{4,5}, S. Puccetti\inst{6,7}, D.\,R.~Ballantyne\inst{8}, F.\,E.\, Bauer\inst{9,10,11,12}, S.\,E.~Boggs\inst{13}, W.\,N.~Brandt\inst{14,15,16}, M. Brightman\inst{3}, F.\,E.~Christensen\inst{17}, W.\,W.~Craig\inst{13,18}, P. Gandhi\inst{19,20}, C.\,J.~Hailey\inst{21}, F.\,A.~Harrison\inst{3}, M.\,J.~Koss\inst{22,26}, G. Madejski\inst{23}, C.~Ricci\inst{9,12}, E. Rivers\inst{3}, D.~Stern\inst{24}, and W.\,W.~Zhang\inst{25}}
   
\institute{INAF-Osservatorio Astronomico di Bologna, via Ranzani 1, 40127 Bologna, Italy\\ \email{\href{mailto:alberto.masini4@unibo.it}{alberto.masini4@unibo.it}}
         \and
Dipartimento di Fisica e Astronomia (DIFA), Universit\`a  di Bologna, viale Berti Pichat 6/2, 40127 Bologna, Italy
         \and
Cahill Center for Astronomy and Astrophysics, California Institute of Technology, Pasadena, CA 91125, USA 	
        \and
New York University Abu Dhabi, P.O. Box 129188, Abu Dhabi, UAE
	\and
Center for Cosmology and Particle Physics, Department of Physics, New York University - Affiliate Member, 4 Washington Place, New York, NY 10003, USA
	\and
ASDC-ASI, Via del Politecnico, 00133 Roma, Italy
	\and
INAF-Osservatorio Astronomico di Roma, Via Frascati 33, 00040 Monte Porzio Catone, Italy
	\and
Center for Relativistic Astrophysics, School of Physics, Georgia Institute of Technology, Atlanta, GA 30332, USA
	\and
Instituto de Astrof\'isica, Facultad de F\'isica, Pontificia Universidad Cat\'olica de Chile, 306, Santiago 22, Chile
	\and
Millennium Institute of Astrophysics, MAS, Nuncio Monse\~nor S\'otero Sanz 100, Providencia, Santiago de Chile
	\and
Space Science Institute, 4750 Walnut Street, Suite 205, Boulder, Colorado 80301
	\and
EMBIGGEN Anillo, Concepci\'on, Chile
	\and
Space Science Laboratory, University of California, Berkeley, CA 94720, USA
	\and
Department of Astronomy and Astrophysics, 525 Davey Lab, The Pennsylvania State University, University Park, PA 16802, USA
	\and
Institute for Gravitation and the Cosmos, The Pennsylvania State University, University Park, PA 16802, USA
	\and
Department of Physics, 104 Davey Lab, The Pennsylvania State University, University Park, PA 16802, USA
	\and
DTU Space National Space Institute, Technical University of Denmark, Elektrovej 327, 2800 Lyngby, Denmark
	\and
Lawrence Livermore National Laboratory, Livermore, CA 94550, USA
	\and
Centre for Extragalactic Astronomy, Department of Physics, University of Durham, South Road, Durham DH1 3LE, UK
	\and
School of Physics and Astronomy, University of Southampton, Highfield, Southampton SO17 1BJ, UK
	\and
Columbia Astrophysics Laboratory, Columbia University, New York, NY 10027, USA
	\and
Institute   for   Astronomy,   Department   of   Physics,   ETH Zurich,  Wolfgang-Pauli-Strasse  27,  CH-8093  Zurich,  Switzerland
	\and
Kavli Institute for Particle Astrophysics and Cosmology, SLAC National Accelerator Laboratory, Menlo Park, CA 94025, USA
	\and
Jet Propulsion Laboratory, California Institute of Technology, Pasadena, CA 91109, USA
	\and
NASA Goddard Space Flight Center, Greenbelt, MD 20771, USA
	\and
SNSF Ambizione Postdoctoral Fellow \\}


\abstract
   {}
   {Study the connection between the masing disk and obscuring torus in Seyfert 2 galaxies. }
   {We present a uniform X-ray spectral analysis of the high energy properties of 14 nearby megamaser Active Galactic Nuclei observed by \textit{NuSTAR}. We use a simple analytical model to localize the maser disk and understand its connection with the torus by combining \textit {NuSTAR} spectral parameters with available physical quantities from VLBI mapping.}
   {Most of the sources analyzed are heavily obscured, showing a column density in excess of $\sim$ 10$^{23}$ cm$^{-2}$. In particular, 79\% are Compton-thick ($N_{\rm H}$ > 1.5 $\times$ 10$^{24}$ cm$^{-2}$). Using column densities measured by \textit{NuSTAR}, with the assumption that the torus is the extension of the maser disk, and further assuming a reasonable density profile, the torus dimensions can be predicted. They are found to be consistent with mid-IR interferometry parsec-scale observations of Circinus and NGC 1068. In this picture, the maser disk is intimately connected to the inner part of the torus. It is probably made of a large number of molecular clouds connecting the torus and the outer part of the accretion disk, giving rise to a thin disk rotating in most cases in Keplerian or sub-Keplerian motion. This toy model explains the established close connection between water megamaser emission and nuclear obscuration as a geometric effect.}
   {}


   \titlerunning{\textit{\textit{NuSTAR}} megamaser AGN}
   \authorrunning{A. Masini et al.} 
   \maketitle
%

\section{Introduction}
There is strong evidence that a significant fraction of Active Galactic Nuclei (AGN) are mildly or heavily obscured by a large amount of gas, preventing us from detecting their nuclear emission. The study of the column density distribution among Seyfert 2 galaxies \citep{1999ApJ...522..157R} is a key element to understand the nature and structure of the putative toroidal reprocessor of the AGN unified model  \citep{1993ARA&A..31..473A, 1995PASP..107..803U}, which is responsible for many of the observed differences between type 1 and type 2 Seyfert galaxies.
The most common way to study the innermost regions of obscured AGN is through their hard X-ray emission, which can penetrate even very high obscuring column densities. \newline The \textit{Nuclear Spectroscopic Telescope Array} (\textit{NuSTAR}) is a recent hard X-ray observatory launched in June 2012. It has two coaligned  X-ray optics which focus X-ray photons onto two independent shielded focal plane modules (FPMs), namely FPMA and FPMB. Thanks to its focusing optics, it has a broad and high quality spectral coverage from 3 to 79 keV, a field of view (FoV) at 10 keV of 10\arcmin~ $\times$ 10\arcmin~ and an 18\arcsec~ FWHM with a half-power diameter of 58\arcsec~ \citep{2013ApJ...770..103H}. Given these features, \textit{NuSTAR} is suitable for studying the hard X-ray spectra of AGN with high sensitivity, discriminating between the transmitted nuclear emission (i.e. radiation which penetrates the obscuring matter along the line of sight) and the scattered or reflected component (i.e. radiation which interacts with circumnuclear gas and gets absorbed or Compton scattered). One of the \textit{NuSTAR} scientific goals is to study the Compton-thick ($N_{\rm H} > 1.5 \times 10^{24}$ cm$^{-2}$) AGN population, which is still poorly understood due to the lack of good quality spectra above 10 keV. This class of obscured active nuclei (see \cite{2004mas..conf..323C}, for a review) is predicted from population synthesis models of the X-ray background \citep{2007A&A...463...79G, 2009ApJ...696..110T} to be a non-negligible contributor to the $\sim$ 30 keV peak of the cosmic X-ray background (CXB), which is still today mostly unresolved  (\cite{2015ApJ...808..185C, 2015ApJ...808..184M}; Aird et al. 2015, in press; Harrison et al. 2015, submitted). \newline The 22 GHz maser line emitted by water vapor molecules having the 6$_{16}$ -- 5$_{23}$ rotational transition can pass through thick absorbing matter and probe the innermost part of the nuclear structure, where high density and near edge-on geometry are needed to produce maser amplification. It has been a long time since the first water vapor extragalactic maser emission was discovered in M33 \citep{1977A&A....54..969C}. Today, nearly 200 galaxies have been detected in H$_2$O maser emission, some associated with disk structures, jets or outflows (e.g., see Table 1 of  \cite{2005ARA&A..43..625L}). Because of their high luminosities with respect to Galactic masers, extragalactic water masers associated to AGN are generally called megamasers, while those associated to star-forming regions are sometimes referred to as kilomasers.
\newline Very long baseline interferometry (VLBI) radio observations provide a robust tool to study subparsec structures. In particular, in disk maser systems, precise estimates of the central dynamical mass can be performed. VLBI maser mapping has been used to test the existence of supermassive black holes (SMBHs) ruling out other candidates (such as rich clusters of stars), and to measure distances independent of a cosmological model \citep[see e.g.]{2009ApJ...695..287R, 2013ApJ...767..154R}. Notably, megamasers are found preferentially in Seyfert 2 (Sy2) galaxies \citep{1997ApJS..110..321B}, and in particular in Compton-thick ones \citep{2008ApJ...686L..13G} which, according to the AGN unification scheme, are likely to be those where the obscuring structure is seen nearly edge-on. 
\newline High quality hard X-ray (> 10 keV) data coupled with high resolution radio maps of the nuclear emission allow new studies of the physics of obscured AGN. Some previous work concentrated on the connection between masing activity and high obscuring column densities in active nuclei, identifying some general and phenomenological results \citep{2008ApJ...686L..13G, 2013MNRAS.436.3388C}. However, many questions are still unanswered. Physical conditions, like the temperature, density and pressure of matter in the vicinity of the SMBH, are still uncertain. It is not completely clear whether the maser emission is associated with the outer part of the accretion disk, or if it is part of the toroidal structure obscuring the nucleus along our line of sight. In this paper we first present new spectral analyses of \textit{NuSTAR} observations of megamaser sources, using a sample of local AGN with good quality X-ray data and radio maps. We then combine the information from the hard X-ray and radio bands to derive a physical picture of the complex environment in which SMBHs are growing. 
\newline The paper is organized as follows: in Section \ref{sec:sample} we present the megamaser sample, and the X-ray analyses with a brief explanation of modeling and some notes on individual sources. Section \ref{sec:results} presents results we obtained combining the spectral and maser disk parameters, with a toy model. A discussion of the toy model is given in Section \ref{sec:discussion}. We give a summary in Section \ref{sec:conclusions}. 
\section{Data and Spectral Analysis}
\label{sec:sample}
\subsection{The Sample}
To build up a sample of disk megamaser sources with high quality maser maps, precise black hole mass estimates and hard X-ray spectral coverage, we cross-correlated a list of VLBI-mapped water megamasers from the Megamaser Cosmology Project\footnote{http://safe.nrao.edu/wiki/bin/view/Main/PublicWaterMaserList} (MCP, see \cite{2012IAUS..287..301H}) with \textit{NuSTAR} observations and well known disk maser sources studied in the literature. We found 11 objects. We then enlarged the sample adding three more sources with VLBI radio maps available, but lacking \textit{NuSTAR} data (refer to \cite{2013MNRAS.436.3388C} for X-ray and maser disk properties of these). The total sample is then composed of 14 sources, which are all the disk water megamasers known today with both precise VLBI maps and hard X-ray spectra. Their main properties are listed in Table \ref{table:sample}. However, we emphasize that this is not a complete sample of all the water megamasers known today, which can be found in \citet{2015ApJ...810...65P}.
\subsection{Data reduction}
We present \textit{NuSTAR} hard X-ray spectral results for 11 sources. In particular, we use archival data for NGC 1194, NGC 1386, NGC 2273, NGC 2960, NGC 3079, NGC 3393, NGC 4388 and IC 2560, for which observation dates and exposure times can be found in Table \ref{table:obs}. For NGC 4945, NGC 1068 and the Circinus galaxy spectral parameters are taken from \citet{2014ApJ...793...26P}, \citet{2014arXiv1411.0670B} and \citet{2014ApJ...791...81A}, respectively. 
\newline The raw events files were processed using the {\it NuSTAR} Data Analysis Software package v. 1.4.1 (NuSTARDAS)\footnote{http://heasarc.gsfc.nasa.gov/docs/nustar/analysis/nustar\_swguide.pdf}. Calibrated and cleaned event files were produced using the calibration files in the {\it NuSTAR} CALDB (20150225) and standard filtering criteria with the \texttt{nupipeline} task.  We used the \texttt{nuproducts} task included in the NuSTARDAS package to extract the {\it NuSTAR} source and background spectra using the appropriate response and ancillary files. We extracted spectra and light curves in each focal plane module using circular apertures of different radii, aimed at optimizing the signal to noise ratio at high energies for every source (see Balokovi\'{c} et al. in prep., for further details). Background spectra were extracted using source-free regions on the same detector as the source. All spectra were binned to a minimum of 20 photons per bin using the HEAsoft task \texttt{grppha}.

\subsection{Spectral Analysis}
The spectral analysis was carried out using the XSPEC software \citep{1996ASPC..101...17A}. We started by fitting the spectra with simple power law models for an initial visual inspection of the broadband spectral curvature and X-ray absorption. We then applied phenomenological models such as \texttt{plcabs} \citep{1997ApJ...479..184Y} or \texttt{pexrav} \citep{1995MNRAS.273..837M} to model the hard X-ray continua. The former describes X-ray transmission of an intrinsic power law with an exponential cutoff through an obscuring medium, taking into account the effects of Compton scattering. The latter models Compton reflection on a slab of neutral material with infinite optical depth. \newline It was always possible to find a combination of these two models which gave an excellent fit to the data. However, as pointed out by \citet{2009MNRAS.397.1549M}, using \texttt{plcabs} and \texttt{pexrav} may produce a bias towards fits dominated by the direct continuum. These initial results, then, need to be tested against more self-consistent and physically motivated models based on Monte Carlo simulations such as \texttt{MY\texttt{Torus}} \citep{2009MNRAS.397.1549M} and \texttt{Torus} \citep{2011MNRAS.413.1206B}. They both model the hard X-ray spectrum emitted through a toroidal reprocessor, consisting of a transmitted continuum (photons passing through the torus without interacting), a scattered or reflected component, made up of photons which interact with matter via Compton scattering, and emission lines (mostly, iron K$\alpha$ and K$\beta$). 
\texttt{MY\texttt{Torus}} allows a dynamic decoupling of these three components to simulate different geometries. It can be used in the default configuration (``\texttt{MYTorus} coupled''), modeling a classical ``donut-shape'' toroidal reprocessor with a fixed covering factor of 0.5 (i.e. the half-opening angle $\theta_{tor}$ of the torus is 60$\degree$, measured as the angle between the axis of the system and the edge of the torus itself), or in a more complex way, called ``\texttt{MYTorus} decoupled''. In this configuration, part of the reflection from the inner far side of the reprocessor could be unobscured by material on the near side of it. In this case, the far-side reflection, at least below $\sim$ 10 keV, can dominate the observed spectrum. This back-reflected continuum and the associated lines are parameterized with a \texttt{MYTorus} face-on reflection spectrum, obtained fixing the inclination angle of the system $\theta_{\rm obs}$ to 0\degree. On the other hand, the forward scattered emission and associated emission lines are approximated using a \texttt{MYTorus} edge-on reflection spectrum, obtained by fixing $\theta_{\rm obs}$ to 90\degree. The relative strength of these two components (front and back-scattered) is encoded in two constants, namely A$_{\rm S90}$ and A$_{\rm S00}$, which are left free to vary. Their respective line components for this geometry are normalized with A$_{\rm L90}$ and A$_{\rm L00}$. Finally, in the most general case, the column density $N_{\rm H}$ obscuring the direct continuum can be decoupled from the column density responsible for the back-reflection or the forward-reflection or both.  For the sake of simplicity, a single $N_{\rm H}$ value is adopted. Refer to \citet{2012MNRAS.423.3360Y} for an exhaustive example on the usage of the model in its decoupled mode. The \texttt{Torus} model does not decouple the three components, but has the opening angle as a free parameter, allowing the measurement of the covering factor (see \cite{2015ApJ...805...41B}). In the following, we always assume a nearly edge-on inclination of the reprocessor (i.e., we fix the inclination angle of the system, $\theta_{\rm obs}$, to ~ 90\degree), even if the toroidal geometry is slightly different between the two models. \newline We first apply all these physical models alone; then, we concentrate on the best among them and refine it adding other line features, if needed, or another power law. The latter is usually significant at lower energies (below $\sim$ 5 keV), and is thought to be due to electron scattering in an ionized zone extended on a size scale larger than the obscuring structure, even if its contribution depends on modeling details. To model in a simple manner this physical situation, we tied all the power law parameters to the primary one (i.e. photon index, redshift, normalization), and multiplied it by a constant, namely $f_{\rm s}$, which is a free parameter in the fit which quantifies the fraction of the primary power law scattered at low energies. We will refer to this power law as a "scattered power law". We explain in detail the general fitting procedure for NGC 1194, while for the remaining eight sources we summarize the most relevant findings, in particular the most precise measurement of the column density available to date. Errors quoted always refer to 90\% confidence limits for one interesting parameter, if not stated otherwise.
\begin{table*}
\caption{The megamaser sample, global properties and references.}             
\label{table:sample}      
\centering          
\begin{tabular}{l c c c c c c c c c c} 
\hline\hline       
\noalign{\vskip 0.5mm}  
Name  & $z$ & $N_{ \rm H}$ & $\log(L_{2-10})$ & Ref. & $M_{\rm BH}$ & Ref. & $R_{\rm d}$ & Disk size & Ref. \\ \noalign{\vskip 0.5mm} 
	& &  [10$^{24}$ cm$^{-2}$] &  [erg s$^{-1}$] & & [10$^{6}$ M$_{\odot}$] & & [pc] & [pc] & \\\noalign{\vskip 0.5mm} 
(1)	& (2)	&	(3)	&	(4)	&	(5)	&	(6)	&	(7)	&	(8) & (9) & (10) \\
\noalign{\vskip 0.5mm}  
\hline                   
\noalign{\vskip 1mm}  
   NGC 1068 & 0.0038 & $> 5.6$ & 43.34 & Bau14 & 8.0 $\pm$ 0.3 & Lod03 & 0.27 & 0.65 - 1.1 & Gre97\\  \noalign{\vskip 0.5mm}  
   NGC 1194 & 0.0136 & 1.4$^{+ 0.3}_{- 0.2}$ & 42.78 & this work & 65 $\pm$ 3 & Kuo11 & 0.14 & 0.54 - 1.33 & Kuo11\\ \noalign{\vskip 0.5mm}  
   NGC 1386 & 0.0029 & 5 $\pm$ 1 & 41.90 & this work & 1.2$^{+ 1.1}_{- 0.6}$ & McC13\tablefootmark{\dag} & 0.05 & 0.44 - 0.94 & Til08 \\ \noalign{\vskip 0.5mm}  
   NGC 2273 & 0.0061 & $> 7.3$ & 43.11 & this work & 7.5 $\pm$ 0.4 & Kuo11 & 0.20 & 0.034 - 0.20 & this work\tablefootmark{*}  \\ \noalign{\vskip 0.5mm}  
   NGC 2960 & 0.0165 & 0.5$^{+ 0.4}_{- 0.3}$ & 41.41 & this work & 11.6 $\pm$ 0.5 & Kuo11 & 0.03 & 0.13 - 0.37 & Kuo11\\ \noalign{\vskip 0.5mm}  
   NGC 3079 & 0.0037 & 2.5 $\pm$ 0.3 & 42.15 & this work & 2.4$^{+ 2.4}_{- 1.2}$ & McC13\tablefootmark{\dag} & 0.07 & 0.4 - 1.3 & Kon05 \\ \noalign{\vskip 0.5mm}  
   NGC 3393 & 0.0125 & 2.2$^{+ 0.4}_{- 0.2}$ & 43.30 & this work & 31 $\pm$ 2 & Kon08 & 0.25 & 0.17 - 1.5 & Kon08 \\ \noalign{\vskip 0.5mm}  
   NGC 4388 & 0.0084 & 0.44 $\pm$ 0.06 & 42.59 & this work & 8.5 $\pm$ 0.2 & Kuo11 & 0.11 & 0.24 - 0.29 & Kuo11 \\ \noalign{\vskip 0.5mm}  
   NGC 4945 & 0.0019 & 3.5 $\pm$ 0.2 & 42.52 & Puc14 & 1.4$^{+ 0.7}_{- 0.5}$ & McC13\tablefootmark{\dag} & 0.10 & 0.13 - 0.41 & this work\tablefootmark{**} \\ \noalign{\vskip 0.5mm}  
   IC 2560	& 0.0098 & $> 6.7$ & 42.98 & this work & 3.5 $\pm$ 0.5 & Yam12 & 0.17 & 0.087 - 0.335 & Yam12 \\ \noalign{\vskip 0.5mm}  
   Circinus	& 0.0015 & 8.7 $\pm$ 1.5 & 42.57 & Are14 & 1.7 $\pm$ 0.3 & Gre03 & 0.11 & 0.11 - 0.4 & Gre03 \\ \noalign{\vskip 0.5mm}  
 \hline  \hline \noalign{\vskip 1mm} 
   NGC 4258 & 0.0015 & 0.087 $\pm$ 0.003 & 41.2 & Cas13 & 39 $\pm$ 3 & Til08 & 0.02 & 0.12 - 0.28 & Til08 \\ \noalign{\vskip 0.5mm}  
   NGC 6264 & 0.0340 & $> 1$ & 42.6 & Cas13 & 29.1 $\pm$ 0.4 & Kuo11 & 0.11 & 0.24 - 0.80 & Kuo11 \\ \noalign{\vskip 0.5mm}  
    UGC 3789 & 0.0109 & $> 1$ & 42.3 & Cas13 & 10.4 $\pm$ 0.5 & Kuo11 & 0.08 & 0.084 - 0.30 & Kuo11 \\
\noalign{\vskip 1mm}    
\hline              
\end{tabular}
\tablefoot{Principal properties of the disk maser AGN sample used in this work. The last three sources are the ones lacking \textit{NuSTAR} data.(1) - Galaxy name. (2) - Redshift. (3) - Best fit intrinsic column density. (4) - Logarithm of the best fit intrinsic (deabsorbed) 2-10 keV luminosity. (5) - References for columns (3) - (4): Are14 - \citet{2014ApJ...791...81A}; Bau14 - \citet{2014arXiv1411.0670B}; Cas13 - \citet{2013MNRAS.436.3388C}; Puc14 - \citet{2014ApJ...793...26P}. (6) - AGN central mass. (7) - References for column (6): Gre03 - \citet{2003ApJ...590..162G}; Til08 - \citet{2008ApJ...678..701T}; Kon08 - \citet{2008ApJ...678...87K}; Kuo11 - \citet{2011ApJ...727...20K}; Lod03 - \citet{2003A&A...398..517L}; McC13 - \citet{2013ApJ...764..184M}; Yam12 - \citet{2012PASJ...64..103Y}. (8) - Dust sublimation radius, calculated using the relation from \citet{2009A&A...502..457G}. See \S\ref{sec:results} for details. (9) - Maser disk inner and outer radii. (10) - References for column (9): Gre97 - \citet{1997Ap&SS.248..261G}; Gre03 - \citet{2003ApJ...590..162G}; Kon05 - \citet{2005ApJ...618..618K}; Kon08 - \citet{2008ApJ...678...87K}; Kuo11 - \citet{2011ApJ...727...20K}; Til08 - \citet{2008ApJ...678..701T}; Yam12 - \citet{2012PASJ...64..103Y}. \newline
\tablefoottext{*}{New maser disk extension estimate from VLBI maps.} \newline
\tablefoottext{**}{Adapted from \citet{1997ApJ...481L..23G}.} \newline
\tablefoottext{\dag}{Maser method mass for which in the original paper an uncertainty is not provided. The error given by \citet{2013ApJ...764..184M} is overestimated.} \newline
\tablefoottext{\dag\dag}{Error replaced by the average error of maser method; see \S\ref{sec:N4388}. }
}
\end{table*}
\begin{table}
\caption{\label{table:obs} \textit{NuSTAR} observation details for the eight sources analyzed.}
\centering
\begin{tabular}{lcc}
\hline\hline
\noalign{\vskip 0.5mm}   
Name & Date of observation & Exposure time \\ \noalign{\vskip 0.5mm} 
           &          &  [ks]     \\
\noalign{\vskip 1mm}  
\hline
\noalign{\vskip 1mm}  
NGC 1194 & 2015-Feb-28 & 31  \\ \noalign{\vskip 0.5mm}  
NGC 1386 & 2013-Jul-19 & 21 \\ \noalign{\vskip 0.5mm}  
NGC 2273  & 2014-Mar-23 & 23 \\ \noalign{\vskip 0.5mm}  
NGC 2960 &  2013-May-10 & 21\\ \noalign{\vskip 0.5mm}  
NGC 3079 &  2013-Nov-12 & 21\\ \noalign{\vskip 0.5mm}  
NGC 3393 &  2013-Jan-28 & 15\\ \noalign{\vskip 0.5mm}  
NGC 4388 & 2013-Dec-27 & 21\\ \noalign{\vskip 0.5mm}  
IC 2560 &  2013-Jan-28, 2014-Jul-16 & 73 \\ \noalign{\vskip 0.5mm}  
\noalign{\vskip 1mm}  
\hline
\end{tabular}
\end{table}
\subsubsection{NGC 1194}
\begin{figure*}
\centering
\subfigure[MYT decoupled + \texttt{zpow} + \texttt{zgauss}]{
 \includegraphics[width = 0.39\textwidth]{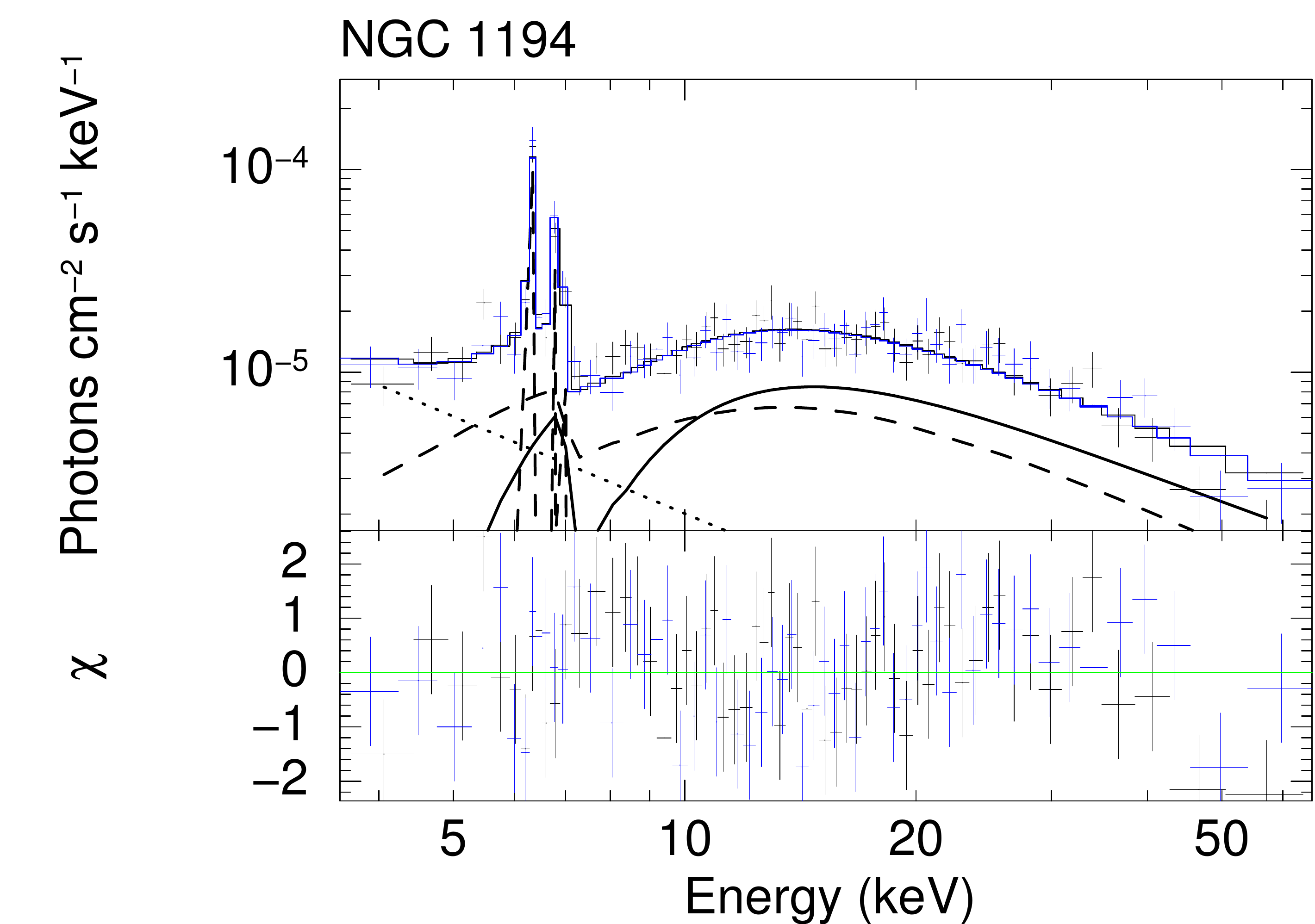}
  \label{fig:subfig1}
  }
\subfigure[Torus + \texttt{zgauss}]{
 \includegraphics[width = 0.39\textwidth]{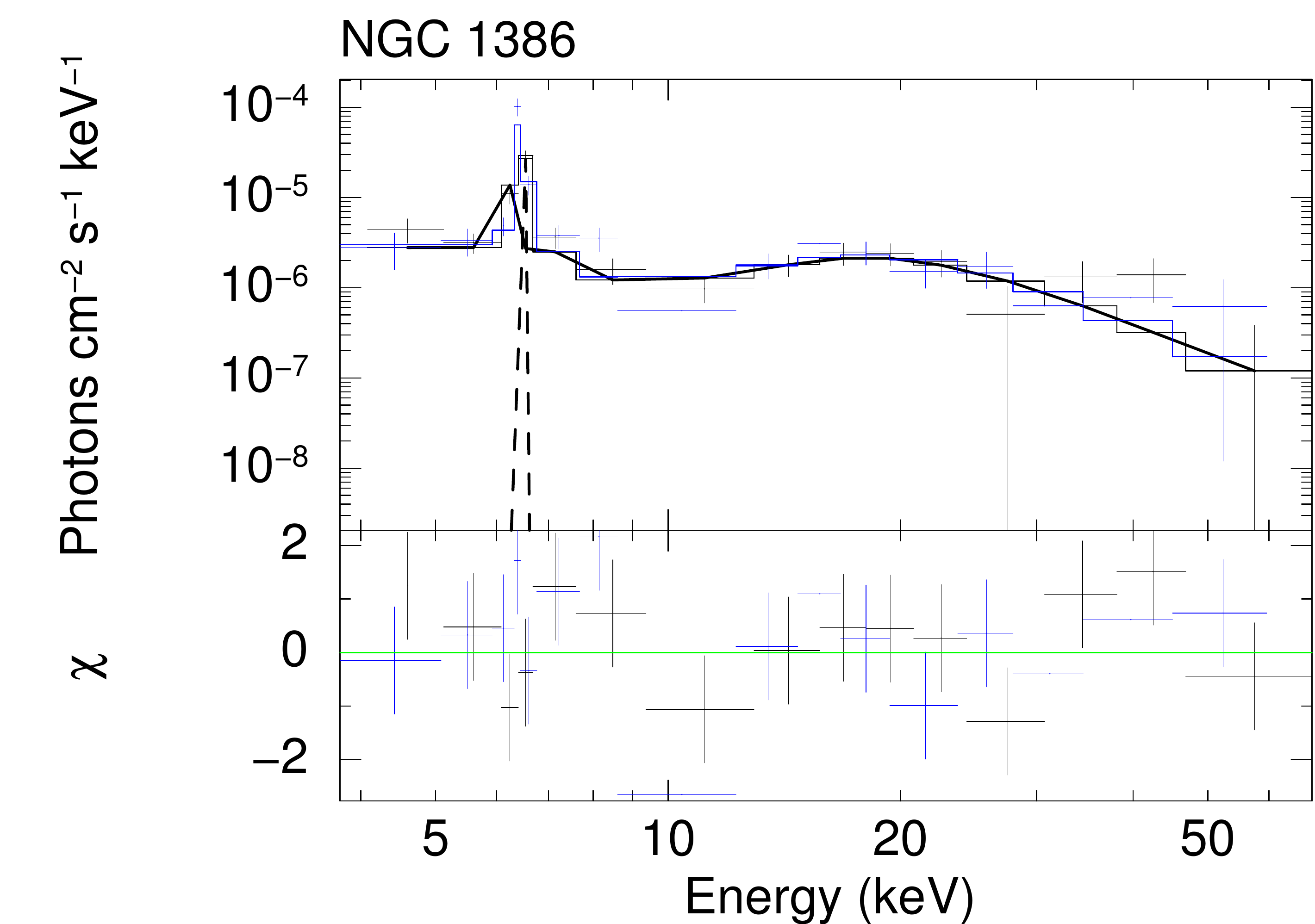}
  \label{fig:subfig2}
  }
  \subfigure[Torus]{
 \includegraphics[width = 0.39\textwidth]{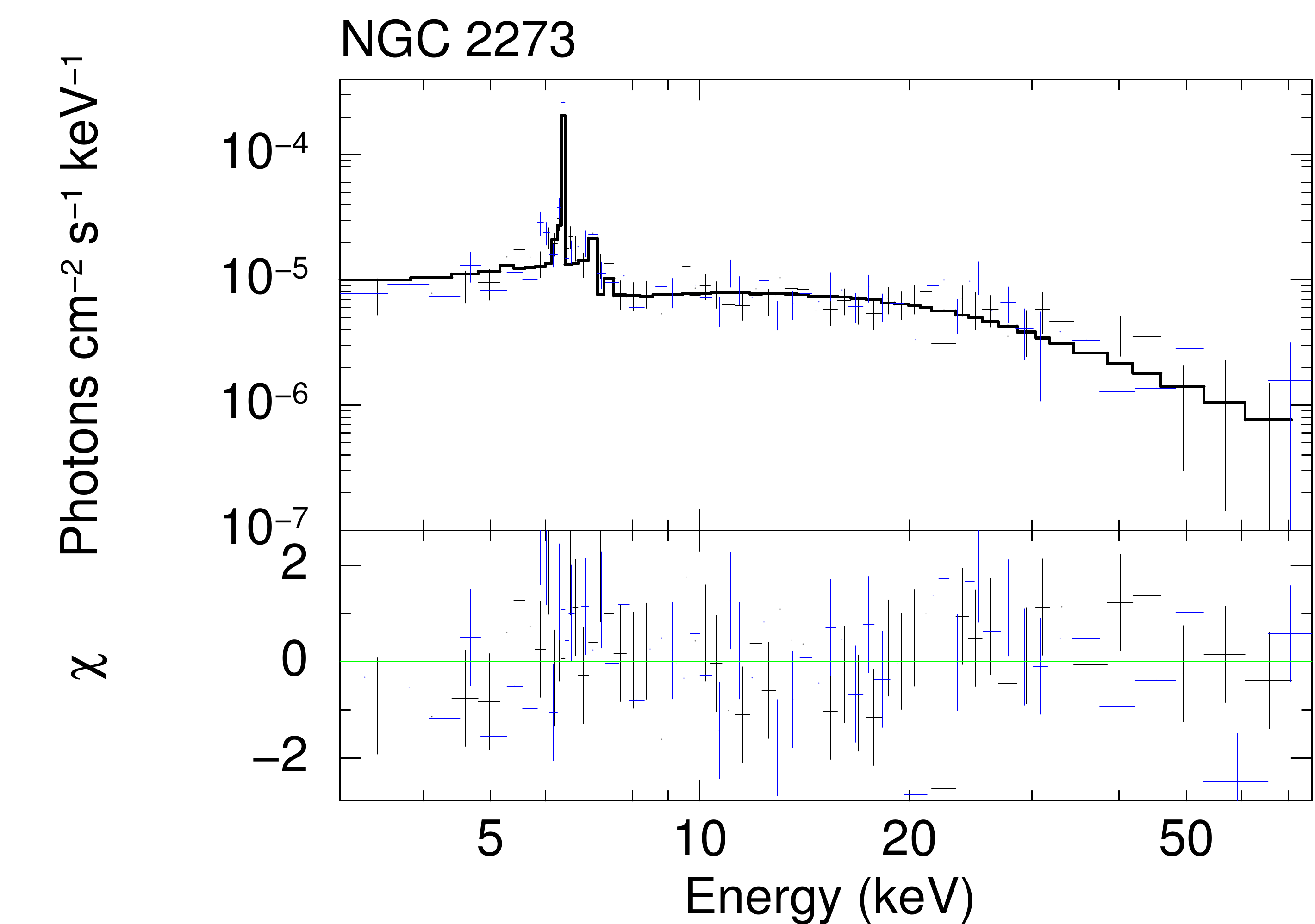}
  \label{fig:subfig3}
  }
\subfigure[MYTorus]{
 \includegraphics[width = 0.39\textwidth]{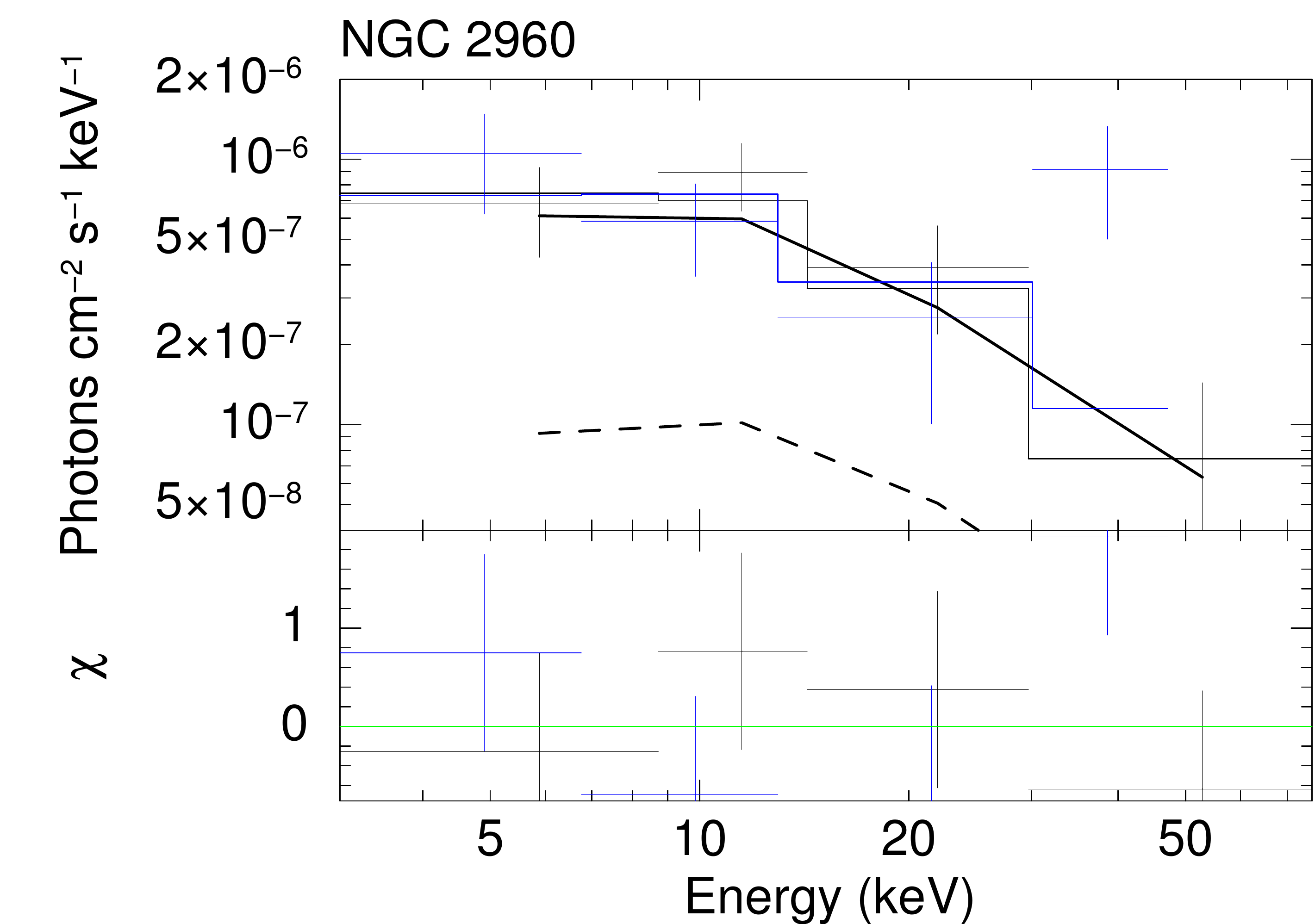}
  \label{fig:subfig4}
  }
  \subfigure[MYTorus + \texttt{zpow}]{
 \includegraphics[width = 0.39\textwidth]{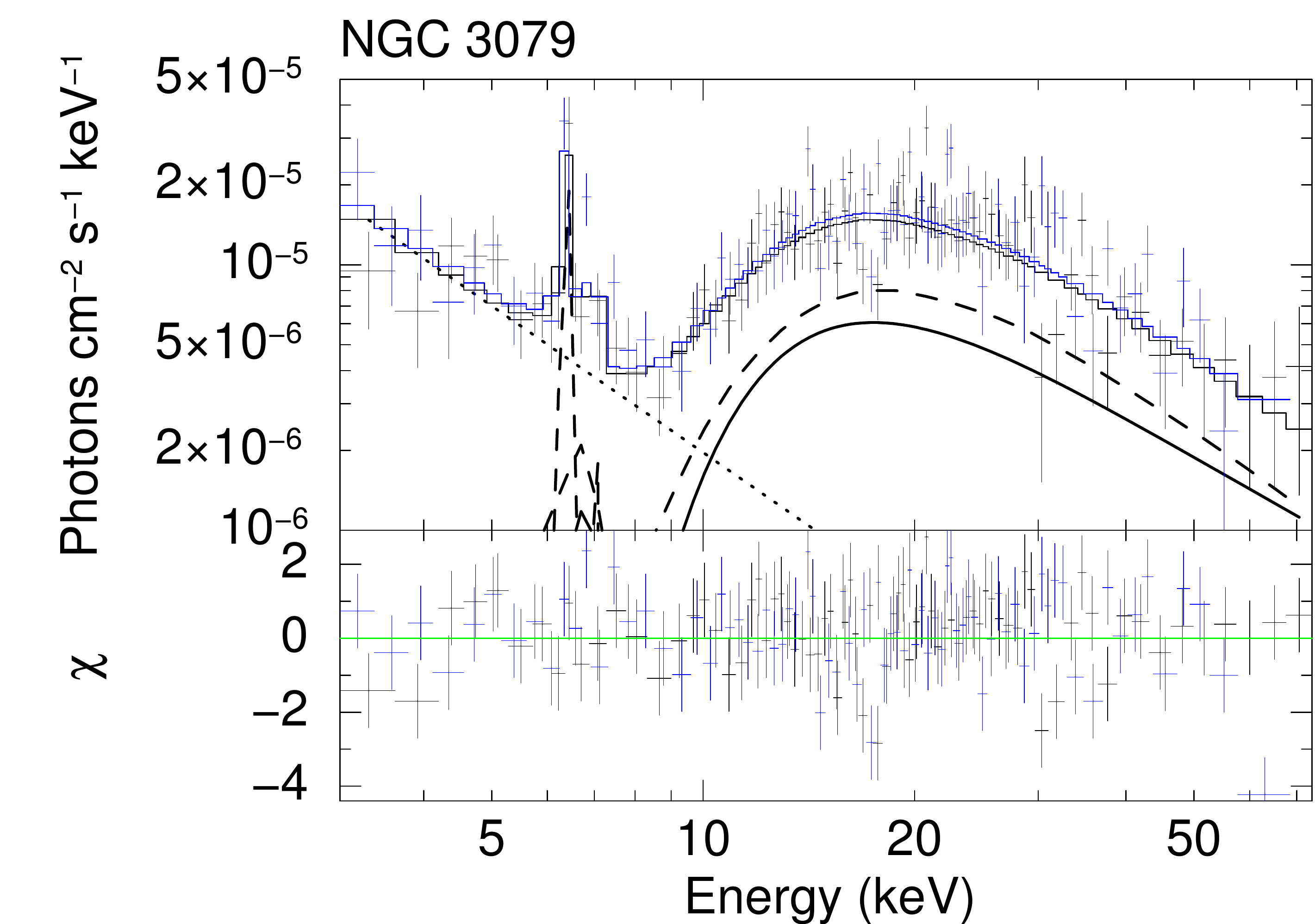}
  \label{fig:subfig5}
  }
   \subfigure[Torus]{
 \includegraphics[width = 0.39\textwidth]{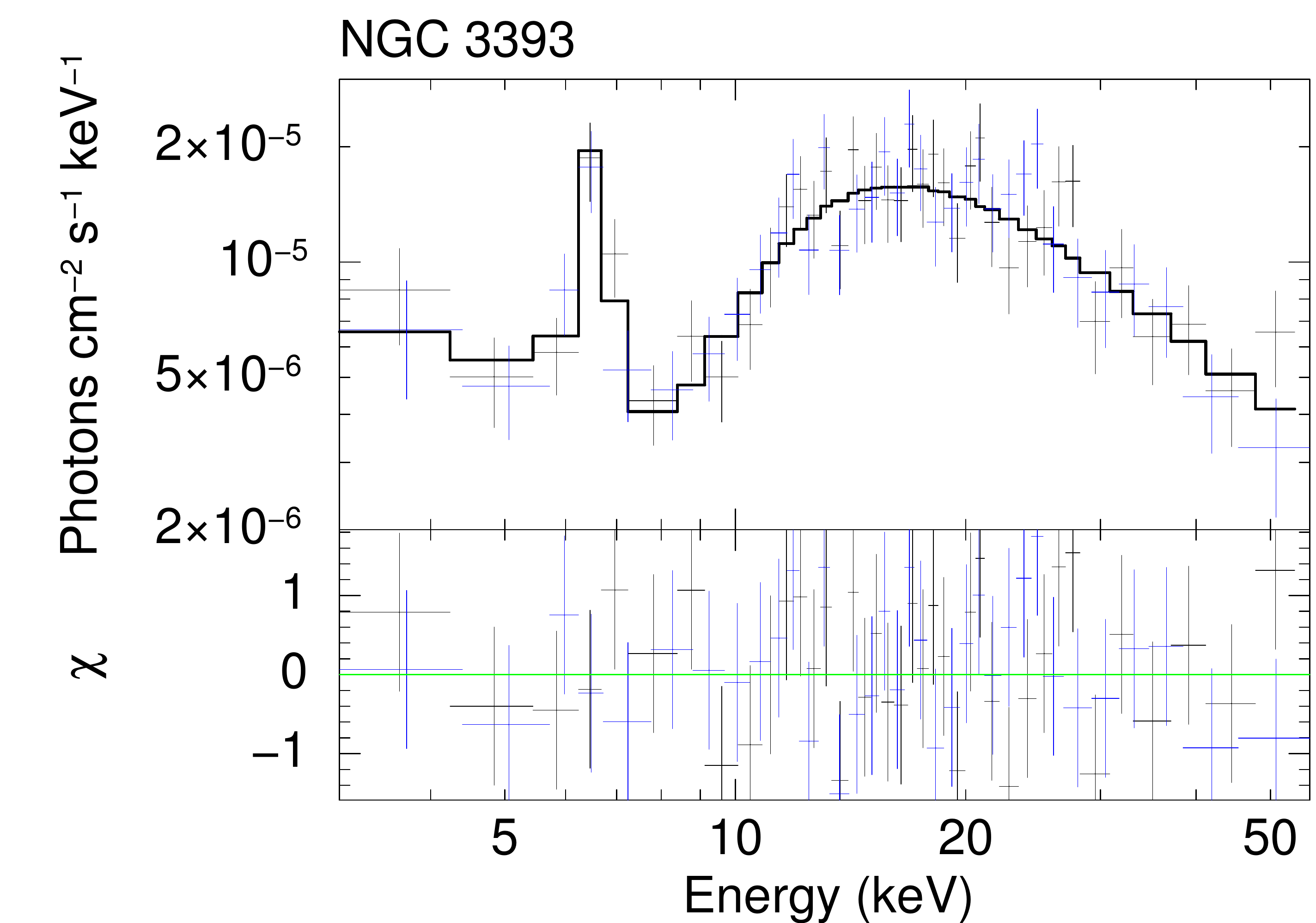}
  \label{fig:subfig6}
  }
   \subfigure[MYT decoupled +  \texttt{zpow}]{
 \includegraphics[width = 0.39\textwidth]{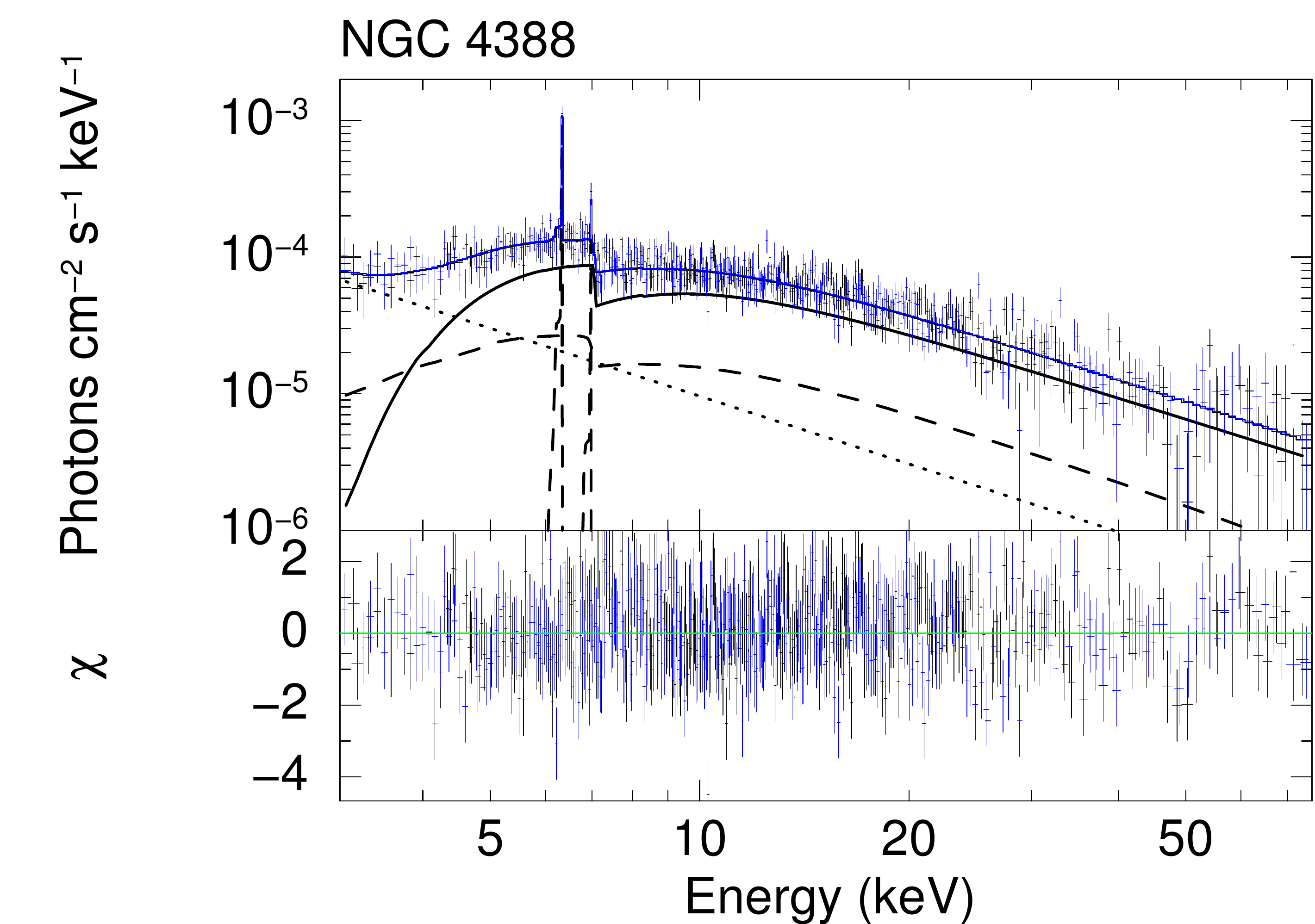}
  \label{fig:subfig7}
  }
\subfigure[Torus + \texttt{zgauss}]{
 \includegraphics[width = 0.39\textwidth]{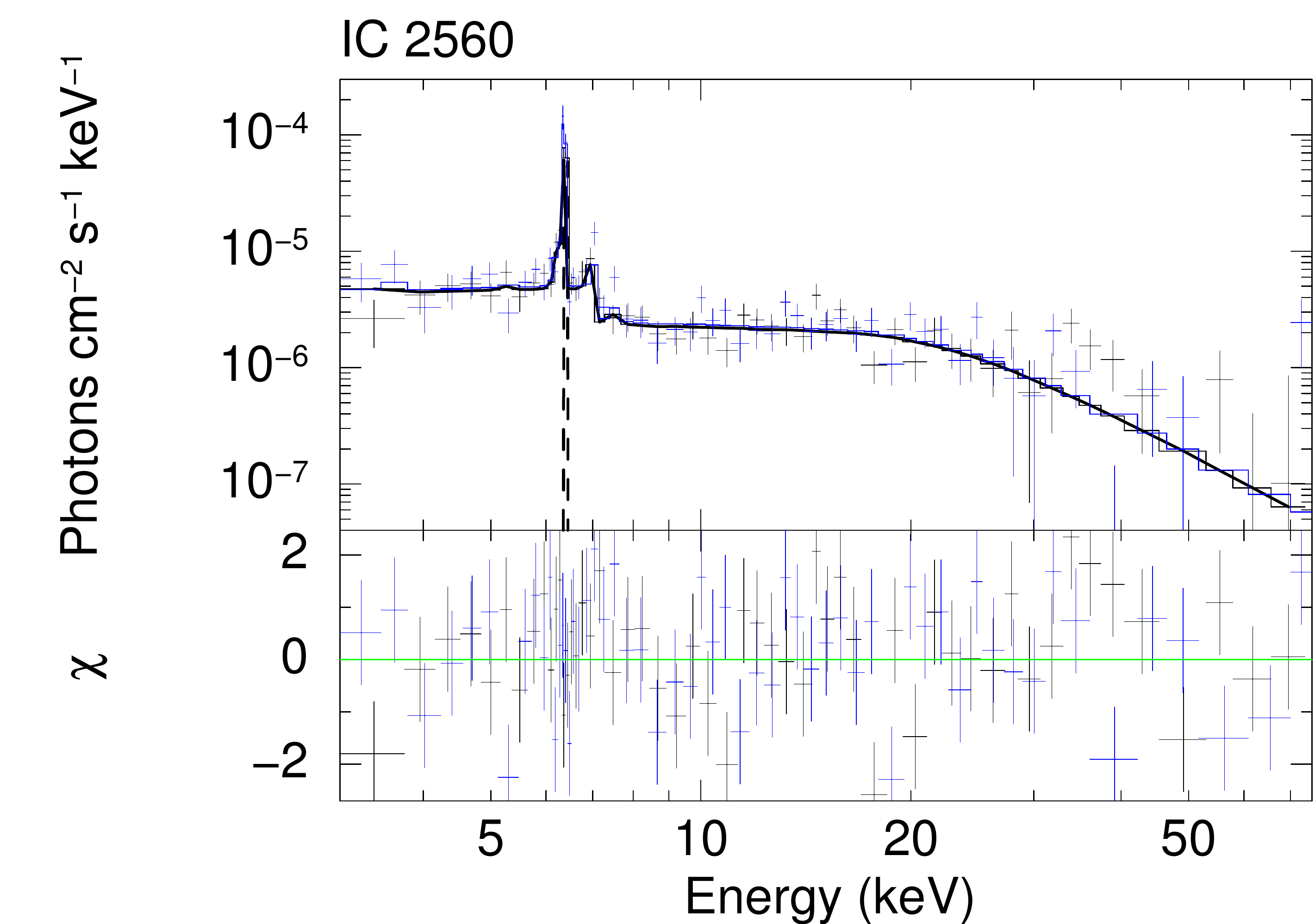}
  \label{fig:subfig8}
  }
\caption{\textit{NuSTAR} spectra, best fitting models and residuals for the eight sources analyzed. FPMA data are shown in black, while FPMB ones are shown in blue. When adopting a \texttt{MYTorus} model, the solid line represents the primary continuum. The reflection and line components are shown as the dashed line, while the scattered power law is shown as the dotted line. When adopting the \texttt{Torus} model, the solid line represents the total spectrum, while add-on line components are shown as the dashed line. } 
\label{fig:subfigure_models}
\end{figure*}

NGC 1194 is a nearby Seyfert 1.9 galaxy. It hosts a circumnuclear maser disk which allowed a precise measurement of the BH mass of (6.5 $\pm$ 0.3) $\times$ 10$^7$ M$_{\odot}$ \citep{2011ApJ...727...20K}.
Fitting the spectrum with an absorbed power law using a Galactic column returns an uncharacteristically hard photon index ($\Gamma$ $\sim$ 0.5) and leaves large residuals, in particular a prominent line feature at $\sim$ 6 - 7 keV and an excess between 10 and 30 keV ($\chi^2$/$\nu$ = 494/117). These are typical spectral signatures of an obscured AGN.  A \texttt{plcabs} model (which accounts for obscuration) with two intrinsically narrow ($\sigma$ = 10 eV) Gaussian components for the lines at 6 - 7 keV returns a much better fit ($\chi^2$/$\nu$ = 186/112). The obscuration is in the Compton-thin regime ($N_{\rm H}$ $\sim$ 6 $\times$ 10$^{23}$ cm$^{-2}$), and the photon index is $\sim$ 1. The residuals still show a hump at $\sim$ 20 keV and signatures of soft excess at energies $<$ 5 keV. A better fit is obtained if \texttt{plcabs} is replaced by a \texttt{pexrav} model ($\chi^2$/$\nu$ = 96/113, $\Gamma$ $\sim$ 1.6). Using both models returns an even better fit, with $\Gamma$ $\sim$ 1.6 and $N_{\rm H}$ $\sim$ 10$^{24}$ cm$^{-2}$ ($\chi^2$/$\nu$ = 85/111), in which the \texttt{plcabs} component is still significant at more than 99\% confidence limit. This appears to be an unphysical situation, since the flux in the reflected component is much larger than the total intrinsic flux of the source. 
\newline  We then apply more physically self-consistent models. An almost edge-on \texttt{Torus} model with a fixed opening angle ($\theta_{tor}$ = 60$\degree$) returns an unacceptable fit ($\chi^2$/$\nu$ = 220/116), with $\Gamma \sim$ 1.4  and $N_{\rm H}$ $\sim$ 7 $\times$ 10$^{23}$ cm$^{-2}$. The fit can be improved by fitting for the  torus opening angle ($\chi^2$/$\nu$ = 201/116), which has a best fit value of $\theta_{tor}$ = 26$\degree$ (the lower limit accepted by the model), with $\Gamma \sim$ 1.4  and $N_{\rm H}$ $\sim$ 6 $\times$ 10$^{23}$ cm$^{-2}$.
A \texttt{MY\texttt{Torus}} model in its default configuration (i.e. coupled mode) returns a fit similar to the \texttt{Torus} one:~$\chi^2$/$\nu$ = 217/116, $\Gamma \sim$ 1.4  and $N_{\rm H}$ $\sim$ 6 $\times$ 10$^{23}$ cm$^{-2}$. A common feature of these models is the underprediction of both the flux of the line component at 6-7 keV, and the emission below 5 keV.
A \texttt{MY\texttt{Torus}} model in its decoupled mode is then applied ($\chi^2$/$\nu$ = 161/115). The front-scattered component vanishes, the photon index is $\sim$ 1.6 and the column density is $\sim$ 10$^{24}$ cm$^{-2}$. These values are consistent with what is found by applying phenomenological models in the first part of the analysis. We note that scattering $\sim$ 3\% of the primary continuum into a scattered power law and adding a line feature at (6.8 $\pm$ 0.1) keV do not change the fundamental fit parameters, but improve it at more than 99\% confidence limit ($\chi^2$/$\nu$ =113/112, $\Delta\chi^2$/$\Delta\nu$ = 48/3).
Even if the best fit model is made up of a combination of \texttt{plcabs}, \texttt{pexrav} and \texttt{zgauss} models ($\chi^2$/$\nu$ = 85/111), we choose to rely on the best fit among the self-consistent ones, as for the other sources, which is the decoupled \texttt{MYTorus} model (see Figure \ref{fig:subfig1}). \newline Summarizing, given that phenomenological models point towards a highly obscured, reflection dominated source, and that the best fit in a physical and self-consistent model is represented by a back-scattered radiation dominated \texttt{MYTorus} model, we conclude that NGC 1194 is a Compton-thick AGN, with the column density $N_{\rm H}$ = 1.4$^{+ 0.3}_{- 0.2}$ $\times$ 10$^{24}$ cm$^{-2}$, consistent with the one reported by \citet{2008ApJ...686L..13G}. We note that, according to the best fit model, reflection dominates below $\sim$10 keV only.
Best fit spectral parameters are given in Table \ref{table:table}.
\newline
\begin{sidewaystable*}
\caption{\textit{NuSTAR} X-ray best fit spectral results for the eight sources analyzed (see text).}
\label{table:table}
\centering
\begin{tabular}{l c c c c c c c c }
\hline\hline
\noalign{\vskip 0.5mm} 
Parameter & NGC 1194 & NGC 1386 & NGC 2273 & NGC 2960 & NGC 3079 & NGC 3393 & NGC 4388 & IC 2560  \\
\hline
\noalign{\vskip 1mm}
Best fit model & MYT D & T & T & MYT & MYT & T &  MYT D & T \\  \noalign{\vskip 0.5mm} 
$\chi^2$/$\nu$ & 113/112 & 32/25 & 123/113 &  6/5\tablefootmark{a} & 189/152 & 49/69 & 693/681 & 125/104 \\ \noalign{\vskip 0.5mm} 
$\Gamma$ & 1.59 $\pm$ 0.15 & > 2.6 & 2.1 $\pm$ 0.1 & 1.9 (f) & 1.8 $\pm$ 0.2 & 1.8 $\pm$ 0.2 & 1.65 $\pm$ 0.08 & 2.7 $\pm$ 0.1  \\ \noalign{\vskip 0.5mm} 
$N_{\rm H}$ [cm$^{-2}$] & 1.4$^{+ 0.3}_{- 0.2}$ $\times$ 10$^{24}$ & (5 $\pm$ 1) $\times$ 10$^{24}$ & > 7.3 $\times$ 10$^{24}$ & 5$^{+ 4}_{- 3}$  $\times$ 10$^{23}$ & (2.5 $\pm$ 0.3) $\times$ 10$^{24}$ & 2.2$^{+ 0.4}_{- 0.2}$ $\times$ 10$^{24}$ &  4.4 $\pm$ 0.6 $\times$ 10$^{23}$ & > 6.7 $\times$ 10$^{24}$ \\ \noalign{\vskip 0.5mm} 
A$_{\rm Z90}$\tablefootmark{b} & 0.003$^{+ 0.003}_{- 0.002}$ & 0.07 $\pm$ 0.01 &  0.07$^{+ 0.02}_{- 0.01}$ & 1.5$^{+ 2.0}_{- 0.7}$ $\times$ 10$^{-4}$ & 0.014$^{+ 0.013}_{- 0.007}$ & 0.017$^{+ 0.012}_{- 0.008}$ & 0.006$^{+ 0.002}_{- 0.001}$ & 0.046$^{+ 0.016}_{- 0.013}$  \\ \noalign{\vskip 1mm} 
A$_{\rm S90}$\tablefootmark{c} & $\sim$ 0 & - & - & 1 (f) & 1 (f) & - & $\sim$ 0 & - \\ \noalign{\vskip 0.5mm} 
A$_{\rm L90}$ & = A$_{\rm S90}$ & - & - & = A$_{\rm S90}$ & = A$_{\rm S90}$ & - &  = A$_{\rm S90}$ & - \\ \noalign{\vskip 0.5mm} 
A$_{\rm S00}$\tablefootmark{c} & 0.8$^{+ 0.4}_{- 0.3}$ & - & - & - & - & - & 1.7$^{+ 0.6}_{- 0.4}$ & - \\ \noalign{\vskip 0.5mm} 
A$_{\rm L00}$ & = A$_{\rm S00}$ & -  & -  & - & - & - & = A$_{\rm S00}$ & -  \\ \noalign{\vskip0.5mm}
$\theta_{\rm tor}$ & - & < 40 & 73$^{+ 6}_{- 11}$ & - & - & 79 $\pm$ 1& - & 53 $\pm$ 23 \\ \noalign{\vskip0.5mm}
$f_{\rm s}$ [\%] & 3 $\pm$ 1 & < 0.5 & < 0.03 & -  & 1.0$^{+ 0.5}_{- 0.4}$ & < 0.4 & 8 $\pm$ 2 & < 0.1 \\ \noalign{\vskip 0.5mm}
$F_{2-10}$ [erg cm$^{-2}$ s$^{-1}$] & 1.2 $\times$ 10$^{-12}$ & 2.8 $\times$ 10$^{-13}$ & 9.5 $\times$ 10$^{-13}$ & 5.8 $\times$ 10$^{-14}$ & 6.2 $\times$ 10$^{-13}$ & 5.1 $\times$ 10$^{-13}$ & 7.8 $\times$ 10$^{-12}$ & 4.0 $\times$ 10$^{-13}$ \\ \noalign{\vskip 0.5mm} 
$F_{10-40}$ [erg cm$^{-2}$ s$^{-1}$] & 1.1 $\times$ 10$^{-11}$ & 1.4 $\times$ 10$^{-12}$ & 5.3 $\times$ 10$^{-12}$& 3.0 $\times$ 10$^{-13}$ & 1.2 $\times$ 10$^{-11}$ & 1.2 $\times$ 10$^{-11}$ & 3.2 $\times$ 10$^{-11}$ & 1.3 $\times$ 10$^{-12}$ \\ \noalign{\vskip 0.5mm} 
$L^{\rm int}_{2-10}$\tablefootmark{d} [erg s$^{-1}$] & 6.0 $\times$ 10$^{42}$ & 8.0 $\times$ 10$^{41}$ & 1.3 $\times$ 10$^{43}$ & 2.6 $\times$ 10$^{41}$ & 1.4 $\times$ 10$^{42}$ & 2.0 $\times$ 10$^{43}$ & 3.9 $\times$ 10$^{42}$ & 9.5 $\times$ 10$^{42}$  \\ \noalign{\vskip 0.5mm} 
$L^{\rm int}_{10-40}$\tablefootmark{d} [erg s$^{-1}$] & 9.6 $\times$ 10$^{42}$ & 1.5 $\times$ 10$^{41}$ & 9.3 $\times$ 10$^{42}$ & 3.1 $\times$ 10$^{41}$ & 1.6 $\times$ 10$^{42}$ & 2.3 $\times$ 10$^{43}$ & 5.7 $\times$ 10$^{42}$ & 2.9 $\times$ 10$^{42}$ \\ \noalign{\vskip 0.5mm} 
FPMB/FPMA & 0.99 $\pm$ 0.06 & 1.1 $\pm$ 0.2 & 1.0 $\pm$ 0.1 & 0.9$^{+ 0.5}_{- 0.4}$ & 1.06 $\pm$ 0.08 & 1.0 $\pm$ 0.1 & 1.02 $\pm$ 0.03 & 1.0 $\pm$ 0.1  \\ 
\noalign{\vskip 1mm} 
\hline
\noalign{\vskip 1mm} 
$\lambda_{\rm Edd}$\tablefootmark{e} & 0.012 $\pm$ 0.003 & 0.09$^{+ 0.08}_{- 0.05}$ & 0.23 $\pm$ 0.06 & 0.0030$^{+ 0.0008}_{- 0.0008}$ & 0.08$^{+ 0.08}_{- 0.04}$ & 0.09 $\pm$ 0.02 & 0.06 $\pm$ 0.02 & 0.4 $\pm$ 0.1 \\
\noalign{\vskip 1mm}
\hline
\end{tabular}
\tablefoot{T = Torus, MYT = MYTorus, MYT D = MYTorus decoupled.\newline \tablefoottext{a}{Data for NGC 2960 are just a \textit{NuSTAR} 3$\sigma$ detection, so we used the Cash statistic \citep{1979ApJ...228..939C} for fitting purposes. We report the reduced $\chi^2$ just for its straightforward interpretation.}\newline
\tablefoottext{b}{Normalization at 1 keV of the direct power law in units of photons keV$^{-1}$ cm$^{-2}$ s$^{-1}$.}\newline
\tablefoottext{c}{Following \citet{2009MNRAS.397.1549M}, A$_{\rm S90}$, A$_{\rm L90}$, A$_{\rm S00}$, A$_{\rm L00}$ are just multiplicative factors of the respective components, and such, lack physical units. To obtain normalizations at 1 keV in units of photons keV$^{-1}$ cm$^{-2}$ s$^{-1}$, the reader should multiply these factors for the primary normalization, i.e. A$_{\rm Z90}$.}\newline
\tablefoottext{d}{Luminosities are intrinsic, i.e. corrected for absorption.} \newline
\tablefoottext{e}{The Eddington ratio is calculated as $\kappa_{\rm bol}$ $\times$ $L_{2-10}^{\rm int}$/$L_{\rm Edd}$, where $\kappa_{\rm bol}$ = 20 $\pm$ 5 is the bolometric correction, constant in our range of 2-10 keV luminosities.}
}
\end{sidewaystable*}
%
\subsubsection{NGC 1386}
NGC 1386 hosts a water maser source \citep{1997AAS...19110402B}, although it is not clear whether the maser spots trace a rotating thin disk or align in front of an underlying continuum (i.e. are jet masers). This makes the central black hole mass estimate challenging and we adopt the one reported by \citet{2013ApJ...764..184M}\footnote{http://blackhole.berkeley.edu/}, which is 1.2$^{+ 1.1}_{- 0.6}$ $\times$ 10$^6$ M$_\odot$.  A default \texttt{MYTorus} model fails ($\chi^2$/$\nu$ = 131/27), and two statistically indistinguishable sets of parameters are possible: one with a Compton-thin obscuration ($N_{\rm H}$  $\sim$ 6 $\times$ 10$^{23}$ cm$^{-2}$) and $\Gamma$ $\sim$ 1.4, and one with a severely obscured AGN ($N_{\rm H}$  $\sim$ 10$^{25}$ cm$^{-2}$) and $\Gamma \sim$ 2.6. A decoupled \texttt{MYTorus} model points toward a back-scattered radiation dominated spectrum, with the same parameters of the coupled case ($\chi^2$/$\nu$ = 65/25). 
The unacceptable fit ($\chi^2$/$\nu$ = 57/27) of the \texttt{Torus} model arises from an underestimation of the line contribution, and provides the best $\chi^2$ among the physical models. 
Following \citet{2015ApJ...805...41B}, we add a line component to the fit at (6.5 $\pm$ 0.1) keV, and get $\chi^2$/$\nu$ = 32/25.
A scattered power law is not required by the data (Figure \ref{fig:subfig2}).
The central source is then obscured by Compton-thick material of column density $N_{\rm H}$  = (5 $\pm$ 1) $\times$ 10$^{24}$ cm$^{-2}$. Best fit parameters are given in Table \ref{table:table}. We note that our results are in agreement with  \citet{2015ApJ...805...41B}, where they focused on the \texttt{Torus} model for a covering factor estimate.

\subsubsection{NGC 2273}
The mass of the SMBH nested in the barred spiral  galaxy NGC 2273 was measured by \citet{2011ApJ...727...20K} to be M$_{BH}$ = (7.5 $\pm$ 0.4) $\times$ 10$^6$ M$_{\odot}$. \newline
The reported parameters put the maser disk very close to the central engine (0.028 - 0.084 pc). This makes NGC 2273 an outlier in some relations \citep{2013MNRAS.436.3388C}. In order to determine whether it is truly an outlier or whether emission at larger radii is missed by the VLBI observations, we derive radii from the more sensitive single dish spectra taken with the Green Bank Telescope \citep{2011ApJ...727...20K}. To do so, we assume a systemic velocity of 1840 km/s (from NED\footnote{http://ned.ipac.caltech.edu/}), a SMBH mass of 7.5 $\times$ 10$^6$ M$_{\odot}$ \citep{2011ApJ...727...20K} and Keplerian rotation \citep{2011ApJ...727...20K}. From the highest and lowest velocity emission (we require the emission to be at least 5 times the RMS) of the high velocity maser features, we find that the innermost radius is $\sim$ 0.034 pc, consistent with \citet{2011ApJ...727...20K}, but that the outermost radius is $\sim$ 0.2 pc. We adopt these values in the following analysis. \newline
A \texttt{MY\texttt{Torus}} model cannot account for the line emission, and gives an unacceptable fit ($\chi^2$/$\nu$ = 267/113). Its decoupled mode provides an acceptable fit ($\chi^2$/$\nu$ = 131/111), pointing towards a back-scattered, reflection dominated AGN.  However, the best fit is found with the \texttt{Torus} model, where the source is heavily Compton-thick and a lower limit on the column density is found ($N_{\rm H}$ > 7.3 $\times$ 10$^{24}$ cm$^{-2}$), consistently with \citet{2005MNRAS.356..295G} and \citet{2009PASJ...61S.317A}. \texttt{Torus} spectral parameters are found in Table \ref{table:table}, while the best fit model is shown in Figure \ref{fig:subfig3}.

\subsubsection{NGC 2960}
The central black hole mass of the spiral megamaser galaxy NGC 2960 (Mrk 1419) as reported by \citet{2011ApJ...727...20K} is (1.16 $\pm$ 0.05) $\times$ 10$^7$ M$_{\odot}$.
NGC 2960 is very faint, and a 21 ks \textit{NuSTAR} snapshot resulted in poor quality data which prevented us from significantly constraining spectral parameters. The fitting procedure in XSPEC was carried out using the Cash statistic \citep{1979ApJ...228..939C}, but we report the reduced $\chi^2$ for direct comparison with other sources. Using a default \texttt{MYTorus} model and fixing the photon index $\Gamma$ = 1.9 (Figure \ref{fig:subfig4}), the fit returns an obscured, but Compton-thin, source ($N_{\rm H}$ = 5$^{+ 4}_{- 3}$ $\times$ 10$^{23}$ cm$^{-2}$), marginally consistent with \citet{2008ApJ...686L..13G} . Results are listed in Table \ref{table:table}. 

\subsubsection{NGC 3079}
The low ionization nuclear emission-line region (LINER) galaxy NGC 3079 presents a thick, flared, probably star forming and self-gravitating maser disk \citep{2005ApJ...618..618K}. The disk outer radius is indeed beyond the sphere of influence radius of the central mass, which is 2.4$^{+ 2.4}_{- 1.2}$ $\times$ 10$^6$ M$_{\odot}$ \citep{2013ApJ...764..184M}. Either \texttt{Torus} ($\chi^2$/$\nu$ = 193/152) or \texttt{MYTorus} ($\chi^2$/$\nu$ = 189/152), both with a scattered power law dominating below 5 keV, give similar results. Using a decoupled \texttt{MYTorus} model ($\chi^2$/$\nu$ = 189/150), the back-scattered contribution vanishes, confirming that the source is dominated by reflection below 10 keV. We therefore choose the coupled \texttt{MYTorus} model as the best fit (Figure \ref{fig:subfig5}), and conclude that NGC 3079 is transmission dominated with a column density of $N_{\rm H}$ = (2.5 $\pm$ 0.3) $\times$ 10$^{24}$ cm$^{-2}$ (see Table \ref{table:table} for other parameters). This result is in agreement with the one found by \citet{2015ApJ...805...41B} using the \texttt{Torus} model.

\subsubsection{NGC 3393}
The nearby barred galaxy NGC 3393 presents an edge-on maser disk which allowed \citet{2008ApJ...678...87K} to measure the central mass to be (3.1 $\pm$ 0.2) $\times$ 10$^7$ M$_{\odot}$.
An excellent fit is found with a \texttt{Torus} model ($\chi^2$/$\nu$ = 49/69), with parameters reported in Table \ref{table:table}. This fit is formally indistinguishable from a \texttt{MYTorus} model with a scattered power law in the soft part of the spectrum, either coupled or decoupled ($\chi^2$/$\nu$ = 49/69 and $\chi^2$/$\nu$ = 48/67, respectively), but we choose \texttt{Torus} as the best fit because it only requires one component (i.e., the scattered power law is not significant, see Figure \ref{fig:subfig6}). However, the spectral parameters are the same within the uncertainties. We therefore conclude that NGC 3393 hosts a Compton-thick AGN ($N_{\rm H}$ = 2.2$^{+ 0.4}_{- 0.2}$ $\times$ 10$^{24}$ cm$^{-2}$), in agreement with the results of \citet{2015arXiv150503524K}. 
\subsubsection{NGC 4388}
\label{sec:N4388}
The Virgo cluster member NGC 4388 hosts an active SMBH of mass (8.5 $\pm$ 0.2) $\times$ 10$^6$ M$_{\odot}$ \citep{2011ApJ...727...20K}. In their paper, \citet{2011ApJ...727...20K} suggest using this mass value with caution because of the lack of systemic maser activity and the inability to robustly assess the Keplerian motion of the maser spots. Among the self-consistent models (\texttt{Torus}, \texttt{MYTorus} coupled, \texttt{MYTorus} decoupled) the latter one gives the best $\chi^2$, although spectral parameters are consistent among all of them.
A \texttt{Torus} model points toward smaller opening angles (i.e. larger covering factor, $\sim$ 0.9) which could account for the line emission ($\chi^2$/$\nu$ = 761/684). Fitting with \texttt{MY\texttt{Torus}} in coupled mode underestimates the line feature ($\chi^2$/$\nu$ = 803/684). This could be due to supersolar iron abundance, a broad range of $N_{\rm H}$ with different covering factors, or to a covering factor greater than that of the model, as suggested by the \texttt{Torus} model. Decoupling \texttt{MYTorus} does not change the general results, while the back-scattered radiation seems to be favoured over the vanishing front-scattered one ($\chi^2$/$\nu$ = 729/682). The fit can be improved adding a scattered power law below 5 keV, which brings the reduced chi-squared to $\chi^2$/$\nu$ = 693/681 (Figure \ref{fig:subfig7}).
According to the best fit model, NGC 4388 is a Compton-thin ($N_{\rm H}$ = 4.2 $\pm$ 0.5 $\times$ 10$^{23}$ cm$^{-2}$) transmission dominated source, and our results agree with the constraints implied by its known hard X-ray variability on scales of days \citep{2012A&A...537A..87C} and months \citep{2011MNRAS.417.1140F}. The \texttt{MY\texttt{Torus}} best fit parameters are reported in Table \ref{table:table}.
\subsubsection{IC 2560}
The barred spiral galaxy IC 2560 hosts a (3.5 $\pm$ 0.5) $\times$ 10$^6$ M$_{\odot}$ active SMBH surrounded by a thin molecular maser disk \citep{2012PASJ...64..103Y}, with an uncertain geometry \citep{2008ApJ...678..701T}. The hard X-ray spectrum is well known from previous studies to be reflection dominated \citep{2014ApJ...794..111B, 2015ApJ...805...41B}.
A default \texttt{MYTorus} model cannot reproduce a reflection dominated spectrum ($\chi^2$/$\nu$ = 364/107). A decoupled version of this model does better ($\chi^2$/$\nu$ = 194/105), where all the radiation is back-scattered by Compton-thick material. 
The best fit is obtained with the \texttt{Torus} model ($\chi^2$/$\nu$ = 172/107), and adding a line component at (6.49 $\pm$ 0.06) keV significantly improves the fit ($\chi^2$/$\nu$ = 127/105, Figure \ref{fig:subfig8}). The column density is found to be > 6.7 $\times$ 10$^{24}$ cm$^{-2}$.
Best fit parameters are given in Table \ref{table:table}, and are in agreement with previous results in the literature.
%
%
\subsection{Summary of spectral analysis results}
In this section we presented hard X-ray spectral analyses for eight megamaser sources observed by \textit{NuSTAR}. Three quarters turn out to be Compton-thick, while one quarter are Compton-thin. Among the latter, NGC 2960 is a 3$\sigma$ detection in the \textit{NuSTAR} snapshot, while NGC 4388 is a well known variable source, presenting column density variability on the scale of days.
\newline Moreover, we note that out of seven sources showing a line feature (we exclude in this argument NGC 2960, because of its weak detection), two do not have their line component well fitted by self-consistent models (\texttt{Torus} or \texttt{MYTorus}). In both cases (NGC 1386 and IC 2560) the line was underestimated by the models. Finally, we note that using a decoupled \texttt{MYTorus} model in the Compton-thin regime (such as in the case of NGC 4388) should be done with caution, since the scattered components could mimic the transmitted primary continuum. \newline However, in this paper we are primarily interested in a robust estimate of the absorption column density, rather than an exhaustive discussion of the spectral properties of individual sources, which will be presented elsewhere (Balokovi\'c et al. in prep.).
As previously stated, the final sample is completed by adding three \textit{NuSTAR}-observed well known megamasers (NGC 4945, NGC 1068, Circinus), and three other maser disk AGN whose X-ray (\textit{XMM-Newton} based) and maser disk parameters are taken from \citet{2013MNRAS.436.3388C}.
Summarizing, the fraction of Compton-thick AGN in our final sample of local disk megamasers is at least $\sim$ 79\% (11/14), comparable to the values reported in previous studies (76\% -- \cite{2008ApJ...686L..13G}; 86\% -- \cite{2013MNRAS.436.3388C}) and confirming the tight relation between heavy obscuration and disk maser emission.

\section{The connection between the maser disk and the torus}
\label{sec:results}
\begin{figure}[ht!]
\centering
\subfigure[]{
\includegraphics[width = 0.5\textwidth]{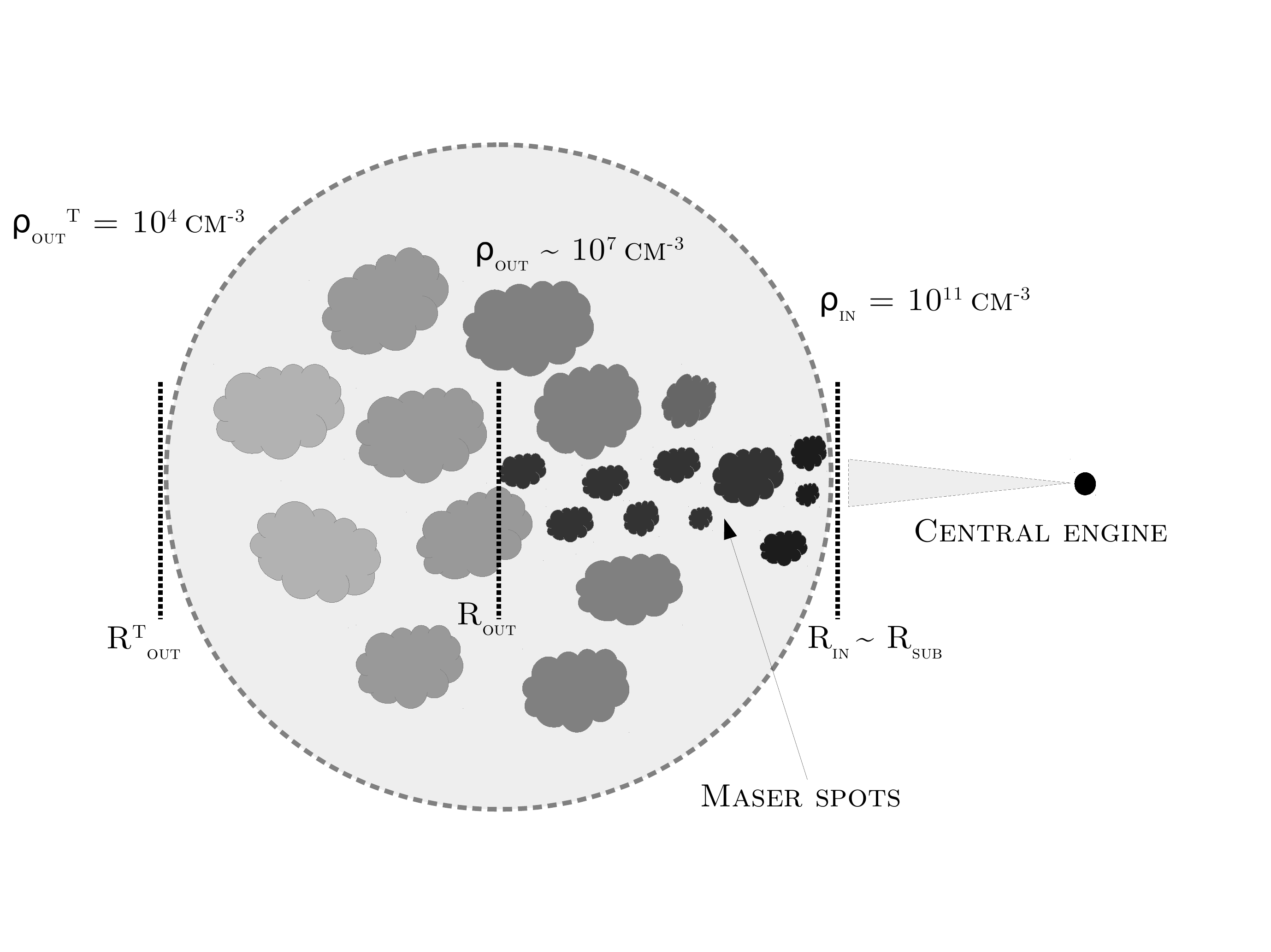}
\label{fig:torus2}
   }
 \subfigure[]{
  \includegraphics[width = 0.5\textwidth]{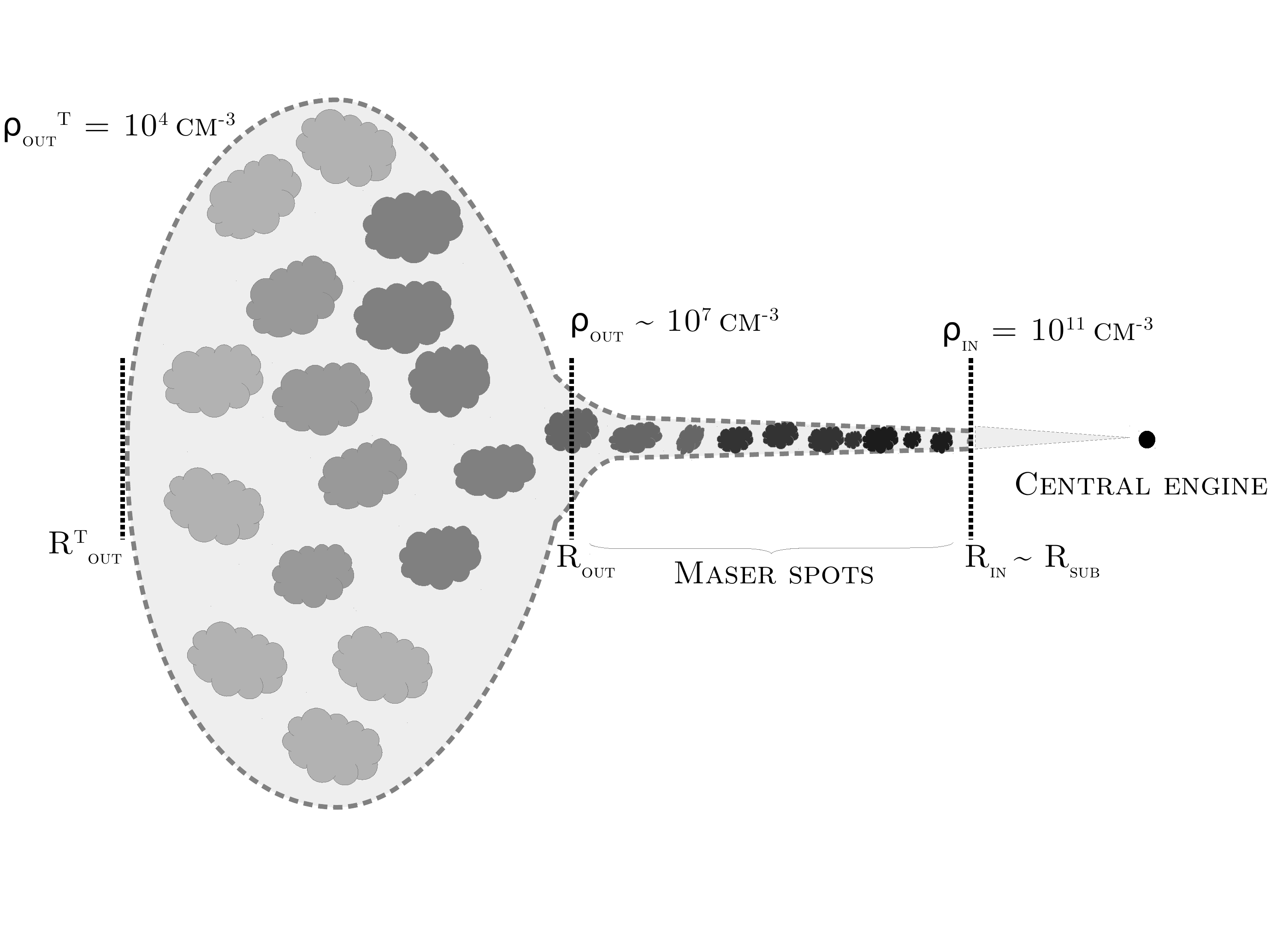}
   \label{fig:torus}
   }
\label{fig:sketch}
\caption{Sketches of the two possible geometries for the location of the maser disk inside the torus. Sizes are not to scale, the sketches are just meant to display the different possible locations of the maser disk. \textit{(a) -} The disk is part of the equatorial plane of the torus. \textit{(b) -} The disk inflates in its outer part, giving rise to a geometrically thicker structure. The change from the inner part to the outer one is not abrupt, and occurs with a gradual change in the dimensions and physical conditions of the clumps, encoded in the same density profile.}
\end{figure}

The aim of this paper is to deepen our understanding of the connection between the torus (seen as the X-ray obscurer) and the maser disk (i.e. an ensemble of clouds orbiting the central black hole, showing water maser activity). First, we can localize the disk. The maser emission occurs too far from the central black hole (in our sample, inner maser radii range from 6.6 $\times$ 10$^4$ to 7.6 $\times$ 10$^6$ gravitational radii) to identify the maser disk with the standard accretion disk, which extends up to $\sim$ 10$^3$ gravitational radii (\cite{2013peag.book.....N}, see \S7.6, pp. 213- 216). Moreover, the presence of water molecules requires the environment to be dusty. We then expect that the maser disk lies outside the dust sublimation radius $R_{\rm d}$, which identifies the torus inner wall within the standard AGN framework \citep[see, e.g.,]{2015arXiv150500811N}. We used the relation from \citet{2009A&A...502..457G} to calculate $R_{\rm d}$ for our sample, adopting a sublimation temperature of 1500 K for graphite grains with an average radial size of 0.05 $\mu$m \citep{1987ApJ...320..537B, 2007A&A...476..713K}. As expected, comparing $R_{\rm d}$ with the inner maser radius $R_{\rm in}$ and considering the uncertainties, all the sources have the maser disk within the dusty zone (i.e., $R_{\rm in} \ge R_{\rm d}$), except NGC 2273 (as already pointed out by \cite{2013MNRAS.436.3388C}). \newline The maser disk can be then generally considered part of the torus, with two different possible geometries: one in which the maser disk is the inner, sub-parsec scale part of the equatorial plane of the classical torus, like in Figure \ref{fig:torus2}, and one in which the masing clouds are tracing a real geometrically thin disk which then inflates into a geometrically thicker end, required to have a large covering factor, as shown in Figure \ref{fig:torus}. 
There are many cases in which the maser disk is seen warped (NGC 2273, NGC 2960, \cite{2011ApJ...727...20K}; NGC 4258, \cite{2005ApJ...629..719H}; NGC 6264, \cite{2011ApJ...727...20K}; Circinus, \cite{2003ApJ...590..162G}), or inflating in its outer part (NGC 3079, \cite{2005ApJ...618..618K}). In particular, in the case of Circinus, the warp is consistent with channeling the nuclear outflow. These geometries are difficult to explain within the framework of Figure \ref{fig:torus2}. Moreover, the basic model of astrophysical maser emission theory predicts that the disk should be directly irradiated by X-rays coming from the central source \citep{1994ApJ...436L.127N}. Again, this makes it difficult to explain the emerging disk when considering a geometry like Figure \ref{fig:torus2}. Finally, a steep density gradient in the vertical direction would be needed to see edge-on maser emission only; otherwise, maser disks would be ubiquitous among Sy2 galaxies, contrary to observations \citep{2011ApJ...742...73Z}. Here we do not have information on the torus vertical structure, but we can exploit the physical properties of the maser emission to infer something about the most likely geometry. High densities, nearly edge-on geometry and a temperature range of $\sim$ 400 -- 1000 K are needed to have maser amplification \citep{2005ARA&A..43..625L}. In this work, we will concentrate only on the density condition. It is indeed difficult to estimate the temperature of the masing gas which is not, by definition, in thermodynamic equilibrium. Instead, we can estimate the density of the masing region, defining the maser disk radial extent $\Delta R$ = $R_{\rm out}$ - $R_{\rm in}$, where $R_{\rm out}$ and $R_{\rm in}$ are the outer and inner radii of the maser disk. They are taken as the locations of the less red/blueshifted and most red/blueshifted maser spots with respect to the systemic velocity of the galaxy, if using the spectrum, or the innermost and outermost spots whether from systemic or red/blueshifted masers, if using the maps.
With the disk extent, and the column density measured from the X-rays, $N_{\rm H}$, we have a rough estimate of the mean density of the material along the line of sight: 
\begin{equation}
\label{eq:rho}
\rho = \frac{N_{\rm H}}{\Delta R} ~ [\text{cm}^{-3}],
\end{equation}
which we can compare with the densities predicted by astrophysical maser theory, 10$^7$ $<$ $\rho$ $<$ 10$^{11}$ cm$^{-3}$ \citep{2005ARA&A..43..625L,2012IAUS..287..323T}. 
Looking at the density distribution in our sample, it is clear that densities obtained with \eqref{eq:rho} are too low, by at least one order of magnitude. This is a hint that using \eqref{eq:rho} and identifying the absorbing medium with the maser disk is not completely appropriate, and if so, all megamaser sources in the sample should be severely obscured. Indeed, with an average density of 10$^9$ cm$^{-3}$ in a fraction of a parsec, the column density of such a maser disk should be of the order of 10$^{26}$ cm$^{-2}$. This is clearly not the case, because 3 out of 14 sources are Compton-thin (i.e. $N_{\rm H}$ $<$ 1.5 $\times$ 10$^{24}$ cm$^{-2}$), and 6 out of 14 are Compton-thick with $N_{\rm H}$ < 10$^{25}$ cm$^{-2}$. \newline In general, the maser disk alone cannot replace the standard torus of the AGN unified model: it is too geometrically thin (otherwise nearly every Sy2 would be identified as a maser source, while nuclear water maser emission detection frequency is low, $\sim$ 3\%, \cite{2011ApJ...742...73Z}), and too optically thick. There could be cases, however, in which a warped disk could simultaneously provide enough obscuration and low covering factor. We will discuss this possibility in \S\ref{sub:warp}. Indeed, five sources of the sample present a lower limit on the column density and are therefore consistent with the absorber being the maser disk itself, seen exactly edge-on.  
Because the density globally increases approaching the black hole, we may simply guess that the maser spots are detected in a high density region, in the inner part of the torus. Moreover, we can explain the tight relation between high obscuration and edge-on maser emission as a co-alignment between the maser clouds and the obscuring matter. In what follows, we then assume geometric alignment and continuity in the radial density profile between the maser disk and the inflated part of the torus, adopting a geometry like Figure \ref{fig:torus}. \newline
An alternative view of the result obtained with Equation \eqref{eq:rho} involves clumpiness, which is quite well addressed by theoretical models. Models like the one by \citet{2006ApJ...648L.101E} study the interplay between the maser disk and the obscuring medium, followed by subsequent relevant work on this topic \citep[e.g.,]{2008ApJ...685..147N, 2008ApJ...685..160N}. Dusty and molecular clouds orbiting the central engine are expected to have column densities in the range $N_{\rm H} \sim 10^{22} - 10^{23}$ cm$^{-2}$, and few clouds are able to provide the necessary obscuration measured with X-ray spectroscopy, together with rapid variability and the radiation reprocessing in the infrared band. Even if many questions are still unanswered, these works point toward the importance of considering a clumpy medium, rather than a smooth one, to interpret and explain many properties of AGN. In this paper, we will use analytical expressions of average quantities, like the density, to get our results. Later on, we shall test whether this methodology is too simplistic or not.
\subsection{A toy model}
Suppose now that the inner and outer radii of the maser disk correspond to the theoretically expected upper and lower limits in density suitable to have maser emission, respectively, and assume a power law for the density profile, such as 
\begin{equation}
\label{eq:rhoprof}
\rho (r) = \rho_{in}\left(\frac{r}{R_{\rm in}}\right)^{- \alpha},
\end{equation}
where $\alpha$ is the power law index which can be estimated for every source taking $\rho_{\rm in}$ = 10$^{11}$ cm$^{-3}$ and $\rho_{\rm out}$ = 10$^{7}$ cm$^{-3}$:
\begin{equation}
\label{eq:alpha}
\alpha = \log \left(\frac{\rho_{\rm in}}{\rho_{\rm out}}\right)/\log \left(\frac{R_{\rm out}}{R_{\rm in}}\right) = \frac{4}{\log \left(\frac{R_{\rm out}}{R_{\rm in}}\right)}.
\end{equation}
Once we have recovered the power law index for each source using the maser disk sizes from Table \ref{table:sample}, a continuity assumption in the radial density profile between the maser disk and the external part of the torus allows us to estimate the torus outer radius. Identifying the outer end of the maser disk with the beginning of the inflated end, and keeping the same $\alpha$, the torus outer radius $R_{\rm out}^{\rm T}$ will be the distance at which the density falls to, say, 10$^4$  cm$^{-3}$ (see \cite{2013peag.book.....N}, \S7.5, pp. 205). This value has a negligible effect on results, as we shall show in the following:
\begin{equation}
\label{eq:torusout}
R_{\rm out}^{\rm T} = R_{\rm out}10^{\frac{1}{\alpha}\log\left(\rho_{\rm out}/\rho_{\rm out}^{\rm T} \right)}=  R_{\rm out}10^{3/\alpha}.
\end{equation}
Assuming density profile continuity between the maser disk and the external part of the torus, with this simple toy model one can recover the torus size, and then integrate its density along the line of sight \textit{inside the inflated part only}, to recover the column density:
\begin{equation}
\label{eq:postnh}
N_{\rm H} = \int_{R_{\rm out}}^{R_{\rm out}^{\rm T}}{\rho(r)dr} = \frac{\rho_{\rm out}R_{\rm out}}{\alpha -1} \left[ 1 - \left(\frac{R_{\rm out}^{\rm T}}{R_{\rm out}}\right)^{1-\alpha}\right],
\end{equation}
\begin{figure}
\centering
\includegraphics[width=0.5\textwidth]{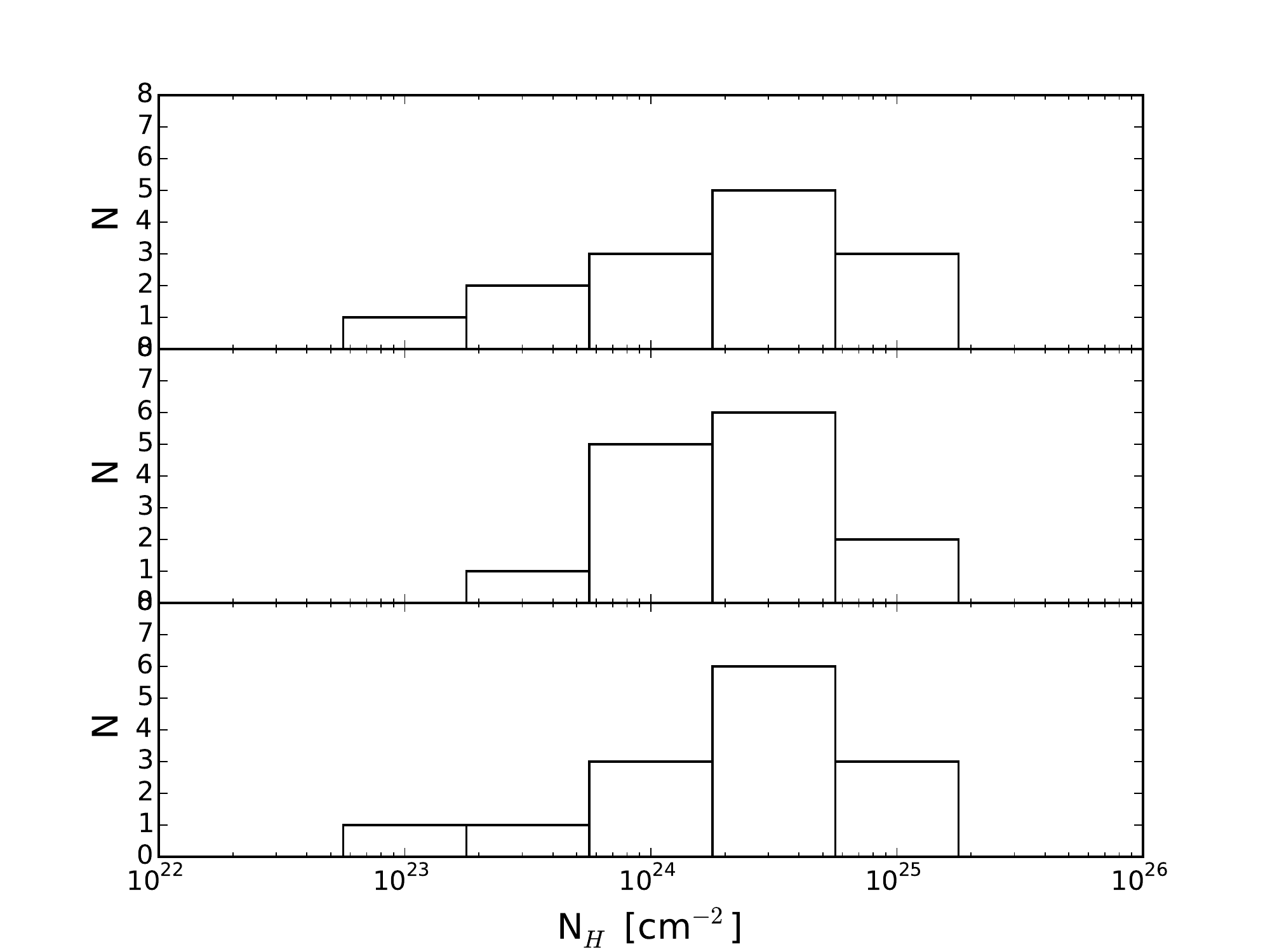}
\caption{\textit{Top panel}: distribution of $N_{\rm H}$ as measured by X-ray spectral fitting. \textit{Middle panel}: distribution of column densities predicted by the model, using a power law density profile. \textit{Bottom panel}: distribution of column densities predicted by the model, with mixed exponential density profile (see text).}
\label{fig:postnh}
\end{figure}
Very interestingly, column densities calculated with \eqref{eq:postnh} are in good agreement with the ones measured with X-ray spectral fitting (Figure \ref{fig:postnh}, middle panel). Moreover, it turns out that between the tunable parameters of the model (the densities $\rho_{\rm in}$ and $\rho_{\rm out}$ at which the maser disk begins and ends, and  $\rho_{\rm out}^{\rm T}$ at which the torus ends), results are sensitive to the outer maser disk density only, $\rho_{\rm out}$. This can be seen directly from Equation \eqref{eq:postnh}. In particular, changing $\rho_{\rm out}$ by one order of magnitude changes $N_{\rm H}$ by a factor $\sim$ 14, while the same variation of $\rho_{\rm in}$ and $\rho_{\rm out}^{\rm T}$ has a negligible impact on the distribution (factor $\sim$ 1.3 and $\sim$ 1.01, respectively).
In other words, a small change of the parameter $\rho_{\rm out}$ gives a large change in the recovered $N_{\rm H}$, and this is a hint that the theoretically driven choice of the three densities is the best at reproducing the observed column density distribution. \newline Another step forward can be made by testing the power law assumption for the density profile. The $\alpha$ parameter in fact only tells how fast the density falls inside the maser disk, decreasing by $\sim$ 4 orders of magnitude in a fraction of a parsec. The resulting distribution of the $\alpha$ parameter is skewed and very steep, peaking at very high values ($\sim$ 8). \newline
Clouds orbiting a SMBH at sub-parsec distances are often modeled with a radial dependence of the form $N(r) \sim r^{-q}$, where $N(r)$ is the number of clouds per unit length, and $q$ is usually 1 or 2 \citep[e.g.,]{2008ApJ...685..160N}. This translates in a radial dependence of the number of clouds per unit volume of $\sim r^{-3q}$. If every cloud has approximately the same number of atoms and same chemical composition, the same radial trend holds also for the density $\rho$ to which we refer here. In Figure \ref{fig:alpha} we show how our $\alpha$ indexes, which describe the density falling rate in a smooth medium, compare with the power law distributions of clouds in clumpy models. In Figure \ref{fig:alpha} we plot the ratio of the outer and inner maser radii as a function of X-ray (deabsorbed) luminosity in the 2-10 keV band. There is no clear correlation between the two quantities. The ratios cluster roughly between the values 2 - 4. We also plot with dashed lines of different colors the ratios expected for different radial distributions, showing that $q$ indexes of 2 - 3 are preferred. One possibility to explain this trend is that the density gradient between the inner and outer maser radii is lower than the assumed one (four orders of magnitude, from $10^{11}$ cm$^{-3}$ at $R_{\rm in}$ to $10^{7}$ cm$^{-3}$ at $R_{\rm out}$). For example, a decrease of three orders of magnitude (which is reasonable, assuming current uncertainties) would make the data fully consistent with a distribution of clouds with power law index $q$ = 2. Another possibility is that in a real medium, which is likely a mixture of clumps, voids and filaments, the density falls abruptly between clouds, steepening the $\alpha$ index. Also warps in maser disks could bias the inner and outer maser radii measurements. Taking into account the above described caveats and uncertainties, we conclude that our analytical formulas for the density are consistent with a radial distribution of clouds $N(r) \sim r^{-q}$, with $q \sim 2,3$. \newline An alternative modelization of our steep density profile is an exponential one:
\begin{figure}[h!]
\centering
\includegraphics[width=0.5\textwidth]{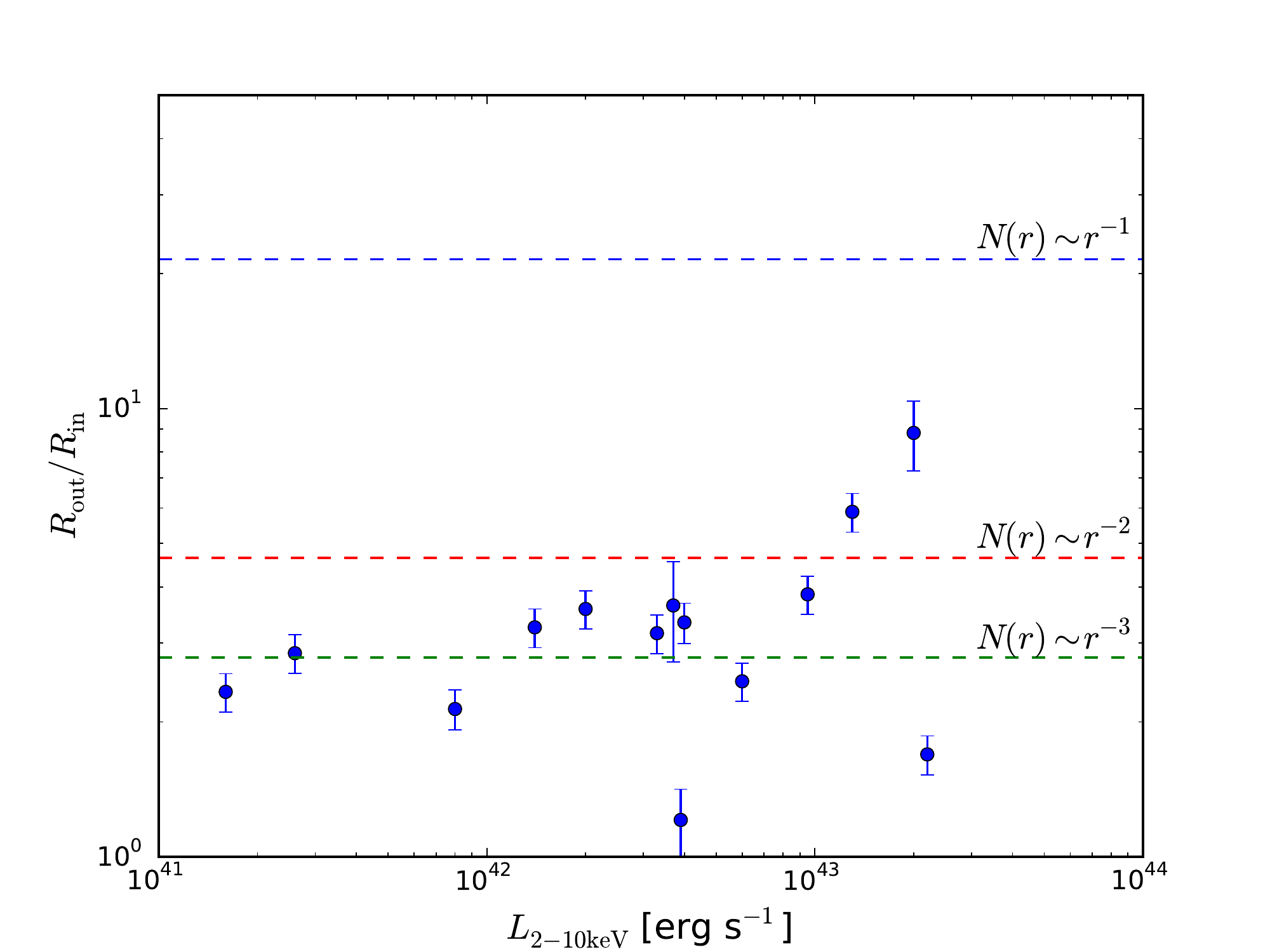}
\caption{Ratio of outer and inner maser radii for the sources in the sample as a function of X-ray (deabsorbed) luminosity in the 2-10 keV band. The trend $N(r) \sim r^{-1}$ (blue dashed line) seems to be ruled out, while $N(r) \sim r^{-2}$ and $N(r) \sim r^{-3}$ (red and green dashed lines, respectively) are preferred. We note that the ratios expected for different clouds distributions depend on the critical maser densities assumed, being the expected ratio for a particular $q$, $R_{\rm out}/R_{\rm in} = 10^{\log(\rho_{\rm in}/\rho_{\rm out})/3q}$. }
\label{fig:alpha}
\end{figure}
\begin{equation}
\label{eq:einasto}
\rho (r) = \rho_{\rm in}\exp\left[-\left(\frac{r-R_{\rm in}}{R_{\rm out}-R_{\rm in}}\right)^{1/n}\ln\left(\frac{\rho_{\rm in}}{\rho_{\rm out}} \right)\right],
\end{equation}
where $n$ is the equivalent of the Sersic index. We note that $n$ = 1 and $n$ = 2 cases are, in general, able to represent Compton-thin and Compton-thick sources, respectively. In other words, results similar to the power law case are found for a mixed density profile, different between Compton-thin and thick sources. \newline For each $n$, the torus outer radius and the column density can be calculated with:
\begin{equation}
\label{eq:torus_rout_exp}
R_{\rm out}^{\rm T} = R_{\rm in} + \left(R_{\rm out} - R_{\rm in}\right)\left[\frac{\ln{(\rho_{\rm out}^{\rm T}/\rho_{\rm in})}}{\ln{(\rho_{\rm out}/\rho_{\rm in})}}  \right]^n 
\end{equation}
\begin{equation}
\label{eq:nhgen}
N_{\rm H} = \frac{n\left(R_{\rm out}-R_{\rm in}\right)}{[\ln{(\rho_{\rm in}/\rho_{\rm out})}]^n}\rho_{\rm in}\left[\Gamma\left(n, \ln{\frac{\rho_{\rm in}}{\rho_{\rm out}}}\right) -  \Gamma\left(n, \ln{\frac{\rho_{\rm in}}{\rho_{\rm out}^{\rm T}}}\right) \right],
\end{equation}
where $\Gamma(n, x)$ is the incomplete Gamma function, and $\Gamma(n, 0)$ = $\Gamma(n)$. In the specific cases $n$ = 1 and $n$ = 2, equation \eqref{eq:nhgen} becomes
\begin{equation}
\label{eq:nhexp}
\small{N_{\rm H} = \begin{cases}
\frac{\left(R_{\rm out} - R_{\rm in}\right)}{\ln{(\rho_{\rm in}/\rho_{\rm out})}}\left(\rho_{\rm out} - \rho_{\rm out}^{\rm T}\right) & n = 1\\ 
\frac{2\left(R_{\rm out} - R_{\rm in}\right)}{[\ln{(\rho_{\rm in}/\rho_{\rm out})}]^2}\left\{\rho_{\rm out}\left[\ln{\left(\frac{\rho_{\rm in}}{\rho_{\rm out}}\right)}+1\right] - \rho_{\rm out}^{\rm T}\left[ \ln{\left(\frac{\rho_{\rm in}}{\rho_{\rm out}^{\rm T}}\right)} +1\right] \right\} & n = 2
\end{cases}.}
\end{equation}
\begin{figure}
\centering
\includegraphics[width=0.5\textwidth]{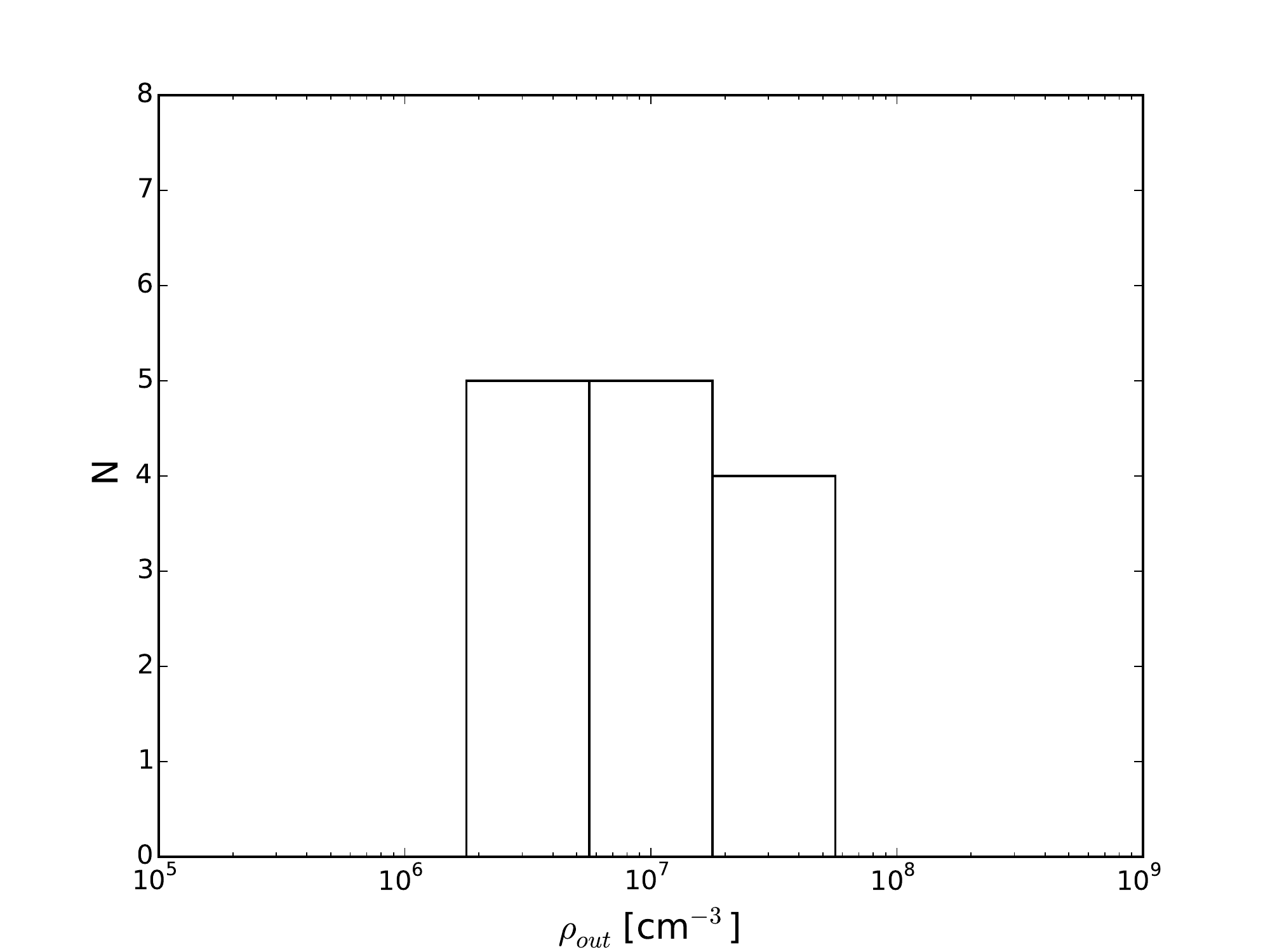}
\caption{Distribution of the parameter $\rho_{\rm out}$ in the sample, calculated as described in \S\ref{torus_size}. See also Table \ref{table:rhoout}. }
\label{fig:rho_out_exp}
\end{figure}
\begin{figure}
\centering
\includegraphics[width=0.5\textwidth]{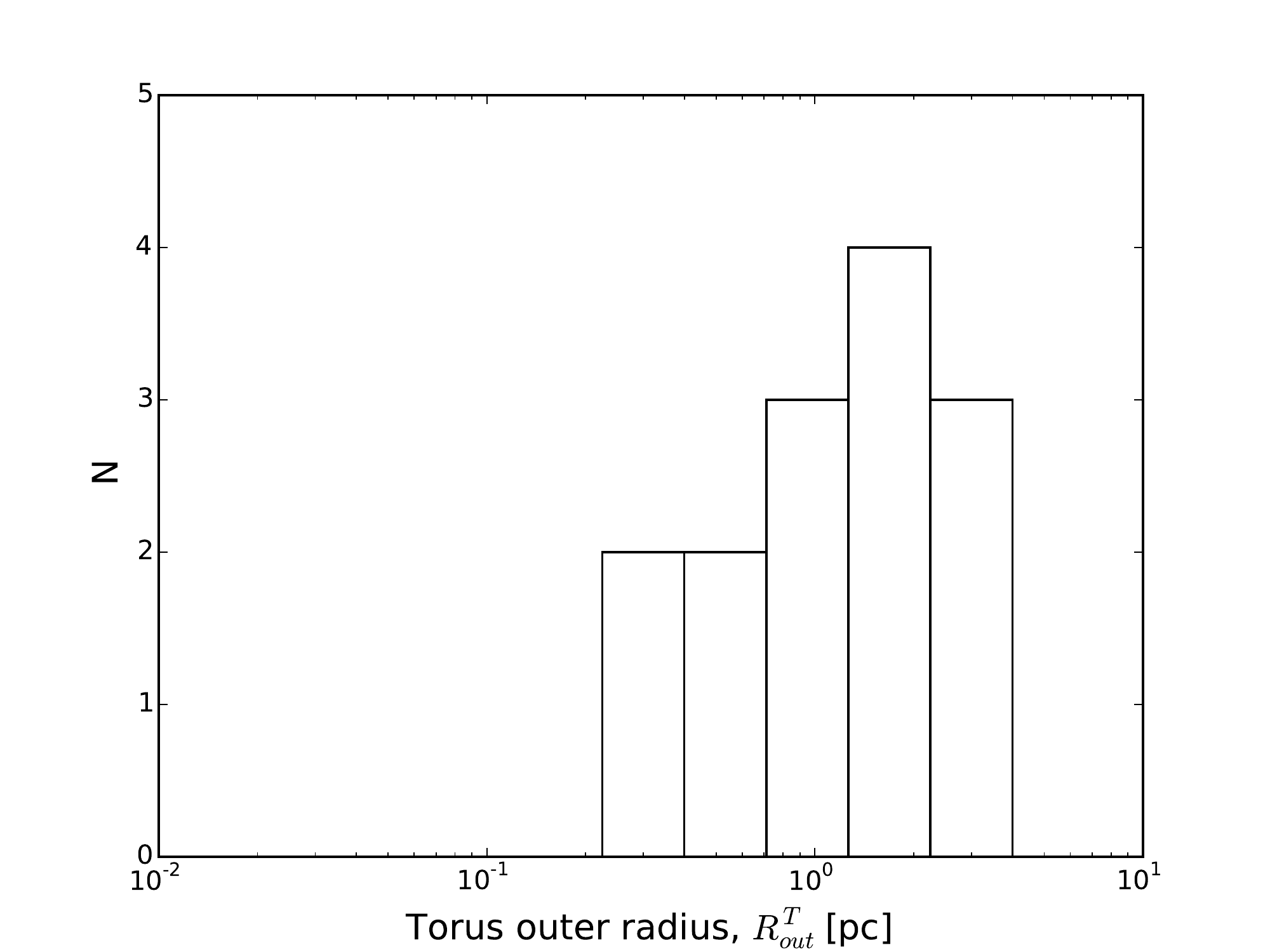}
\caption{Distribution of torus outer radius predicted by the model, calculated as described in \S\ref{torus_size}. See also Table \ref{table:rhoout}.}
\label{fig:size}
\end{figure}
\begin{table}
\caption{$\rho_{\rm out}$ values needed to have $N_{\rm H, meas}$ = $N_{\rm H, pred}$ with the assumption of maser disk-torus density profile continuity, and output of the model, the torus outer radius.}           
\label{table:rhoout}    
\centering                       
\begin{tabular}{l r r} 
\hline\hline
\noalign{\vskip 0.5mm}              
Name & $\rho_{\rm out}$ [cm$^{-3}$] & $R_{\rm out}^{\rm T}$ [pc]\\ 
\noalign{\vskip 0.5mm}    
\hline
\noalign{\vskip 1mm}                          
NGC 1194 & 2.8 $\times$ 10$^{6}$ & 2.4 $\pm$ 0.2 \\ \noalign{\vskip 0.5mm}  
NGC 1386 & 1.3 $\times$ 10$^{7}$ & 2.1 $\pm$ 0.3 \\ \noalign{\vskip 0.5mm}  
NGC 2273 & 4.8 $\times$ 10$^{7}$ & 0.77$^{+ \infty}_{- 0.07}$\\ \noalign{\vskip 0.5mm}  
NGC 2960 & 6.5 $\times$ 10$^{6}$ & 0.53$^{+ 0.06}_{- 0.05}$\\ \noalign{\vskip 0.5mm}
NGC 3079 & 4.1 $\times$ 10$^{6}$ & 2.7 $\pm$ 0.2\\ \noalign{\vskip 0.5mm}
NGC 3393 & 2.6 $\times$ 10$^{6}$ &  3.3 $\pm$ 0.5\\ \noalign{\vskip 0.5mm}
NGC 4388 & 2.4 $\times$ 10$^{7}$ & 0.34 $\pm$ 0.07 \\ \noalign{\vskip 0.5mm}
IC 2560 & 3.1 $\times$ 10$^{7}$ &  1.1$^{+ \infty}_{- 0.1}$\\ \noalign{\vskip 0.5mm}
NGC 1068 & 1.6 $\times$ 10$^{7}$ & 2.2$^{+ \infty}_{- 0.3}$ \\ \noalign{\vskip 0.5mm}
NGC 4945 & 1.6 $\times$ 10$^{7}$ & 1.1 $\pm$ 0.1 \\ \noalign{\vskip 0.5mm}
Circinus & 3.4 $\times$ 10$^{7}$ & 1.3 $\pm$ 0.3 \\ \noalign{\vskip 0.5mm}
NGC 4258 & 1.9 $\times$ 10$^{6}$ & 0.36 $\pm$ 0.03 \\ \noalign{\vskip 0.5mm}
NGC 6264 & 2.8 $\times$ 10$^{6}$ & 1.6$^{+ \infty}_{- 0.2}$ \\ \noalign{\vskip 0.5mm}
UGC 3789 & 6.6 $\times$ 10$^{6}$ & 0.69$^{+ \infty}_{- 0.06}$ \\ 
\noalign{\vskip 1mm}
\hline               
\end{tabular}
\tablefoot{Note that six of our sources have a lower limit on the column density and therefore on the torus outer radius too. In these cases, the derived torus outer radius may be less constrained due to the high obscuring column, compatible with the maser disk one.}
\end{table}
We can then repeat the same analysis, using equations \eqref{eq:torus_rout_exp} and \eqref{eq:nhexp} to predict the column density distribution with fixed $\rho_{\rm out}$ = 10$^7$ cm$^{-3}$ (Figure \ref{fig:postnh}, bottom panel): notably, the mixed exponential density profile (i.e. with 1 $\leq$ $n$ $\leq$ 2) can reproduce the observed distribution of X-ray measured column densities better than the power law general case.
\subsection{The torus size}
\label{torus_size}
Instead of assuming the transition density $\rho_{\rm out}$ to infer $N_{\rm H}$, we now use the column densities measured by \textit{NuSTAR} and reverse the problem. Inverting \eqref{eq:nhexp}, one can calculate the parameter $\rho_{\rm out}$, which is the crucial one, needed to have a torus with a column density equal to the measured one. The result is that it is sufficient to have a sharp distribution of $\rho_{\rm out}$ peaked at 10$^7$ cm$^{-3}$ to have tori with the measured column densities. In other words, fixing the inner maser density $\rho_{\rm in}$ and the outer torus density $\rho_{\rm out}^{\rm T}$ and using the measured column densities, the model points toward a transition density of about 10$^7$ cm$^{-2}$, without knowing anything of the previous theoretical assumptions. \newline Figure \ref{fig:rho_out_exp} shows the resulting $\rho_{\rm out}$ distribution (see Table \ref{table:rhoout} for numerical values), while Figure \ref{fig:size} shows the torus outer radius distribution obtained using equation \eqref{eq:torus_rout_exp} (numerical values are reported in Table \ref{table:rhoout}). 
 
\section{Discussion}
\label{sec:discussion}
\subsection{Comparison with mid-infrared interferometry}
Our toy model allows us to predict the X-ray column density distribution of a sample of disk maser systems, or to calculate the torus outer radius if the column density is known. When dealing with very high column densities (i.e. lower limits on $N_{\rm H}$), the torus outer radius is poorly constrained, since the measured column density can be ascribed to the maser disk without the need of an inflated torus. However, it is interesting to compare our results with mid-infrared (MIR) measurements, which are thought to probe the dusty structure surrounding AGN. In our sample, only NGC 1068 \citep{2009MNRAS.394.1325R, 2014A&A...565A..71L} and Circinus \citep{2007A&A...474..837T, 2014A&A...563A..82T} have been observed with MIR interferometry. In both sources two distinct structures responsible for the MIR emission are detected. One is an elongated, disk-like structure, co-aligned and co-spatial with the maser emission spots and perpendicular to the ionization cones (albeit with the caveat of uncertainty in the absolute astrometry). The second structure, whose origin and theoretical explanation is still unclear, seems to be responsible for diffuse emission on much larger scales (> 1 pc), broadly perpendicular to the first. The geometry of this double dusty structure is currently challenging the classical torus framework. These two-component structures are not considered in our simple toy model; however, we note that the sizes are in broad agreement with our predictions, being parsec-scale. Another possibility of comparison comes from considering half-light radii ($r_{1/2}$), enclosing half of the MIR flux of the source, as done in \citet{2013A&A...558A.149B}. 
We note that the NGC 1068 and Circinus outer radii are less than a factor of two larger than the $r_{1/2}$ values reported by \citet{2013A&A...558A.149B}. This could be expected, since the outer torus radius should be larger than the half-light one. Moreover, the $r_{1/2}$ of NGC 1068 is broadly twice the $r_{1/2}$ of Circinus; the same happens with $R_{\rm out}^{\rm T}$ in our toy model. Future observations in the MIR band of other sources are needed to probe this scenario.
\subsection{Trend with bolometric luminosity}
\begin{figure}[ht!]
\centering
\subfigure[]{
\includegraphics[width = 0.43\textwidth]{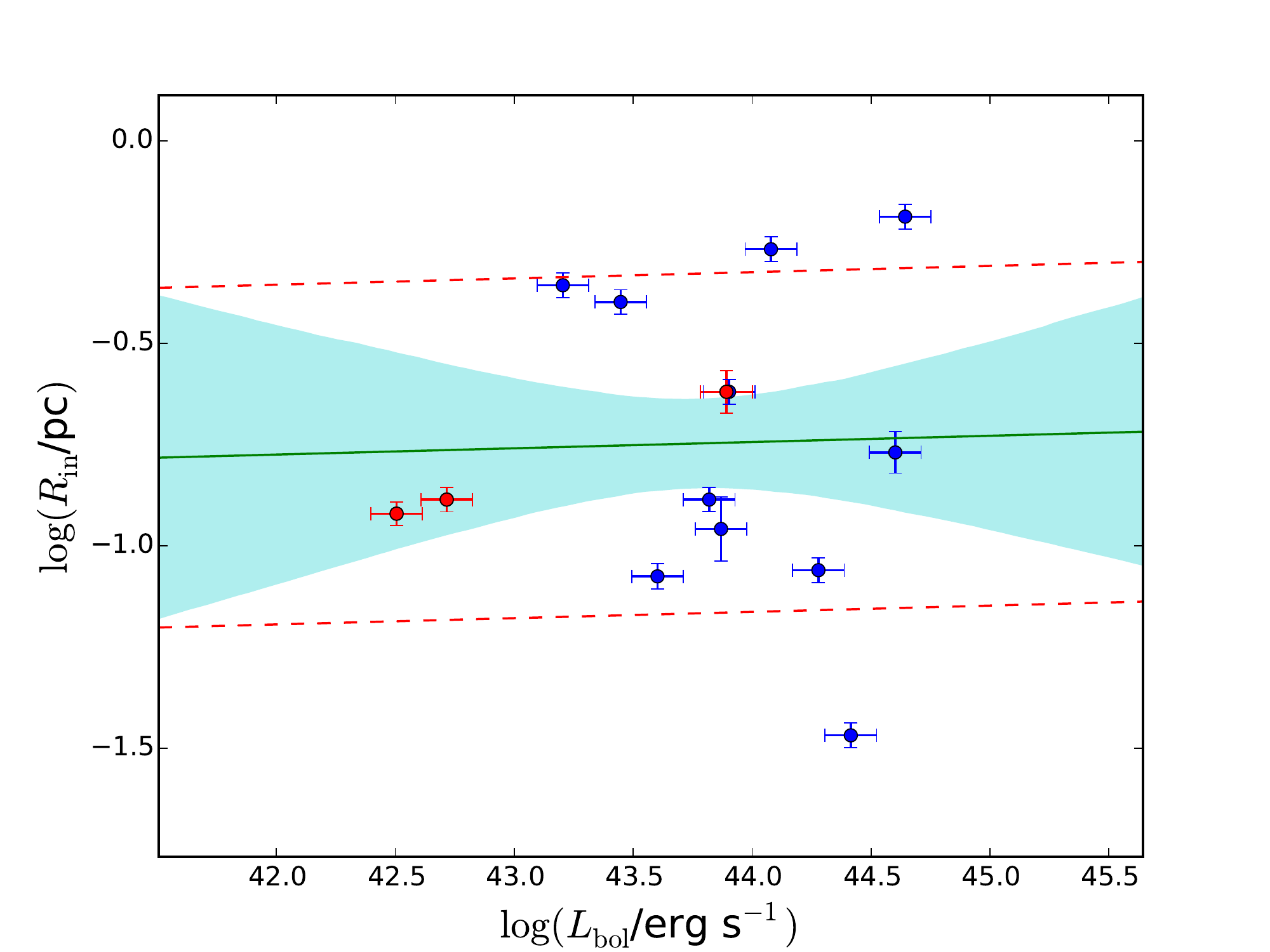}
\label{fig:rin_lbol}
   }
 \subfigure[]{
  \includegraphics[width = 0.43\textwidth]{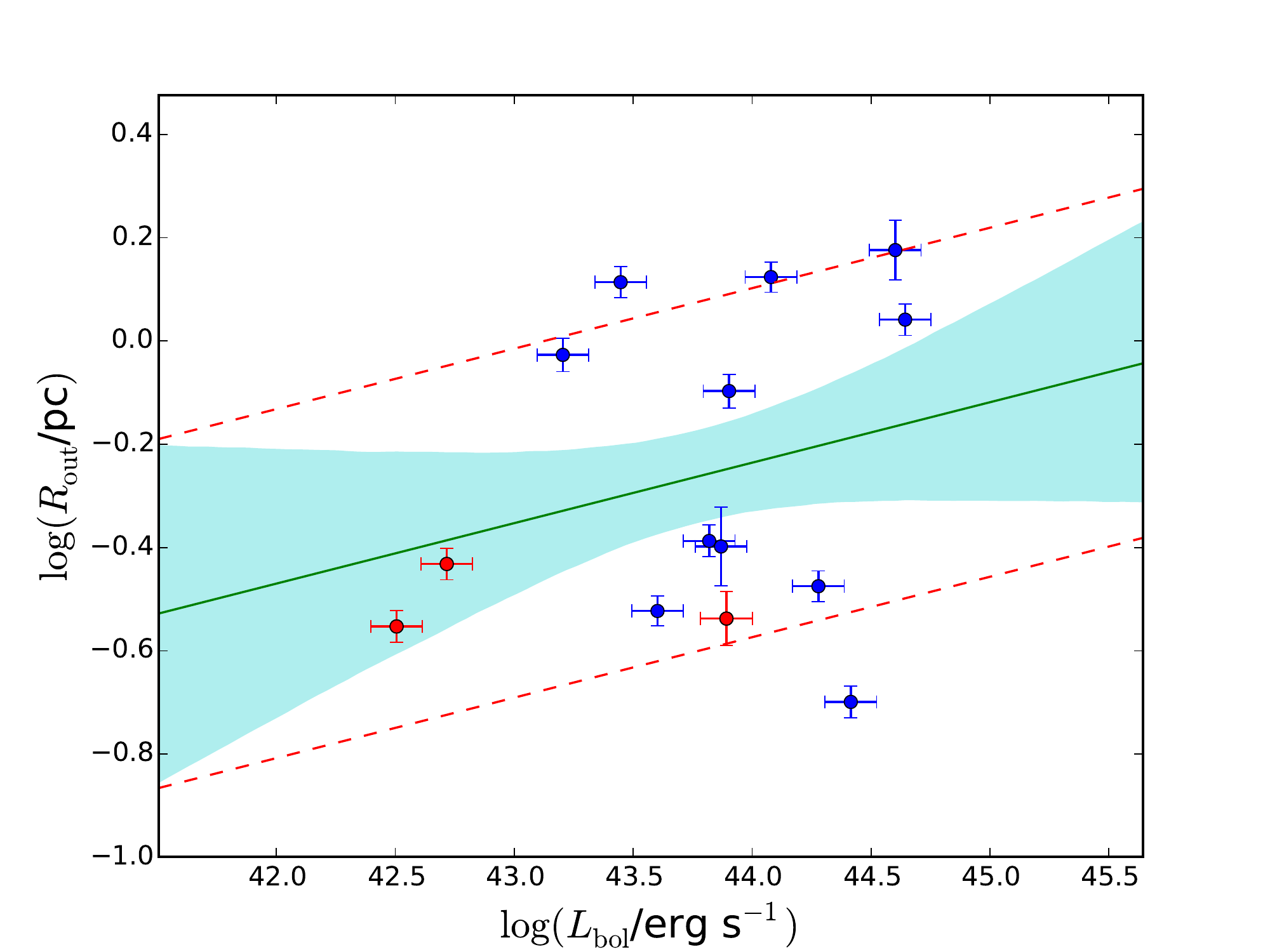}
   \label{fig:rout_lbol}
   }
   \subfigure[]{
\includegraphics[width = 0.43\textwidth]{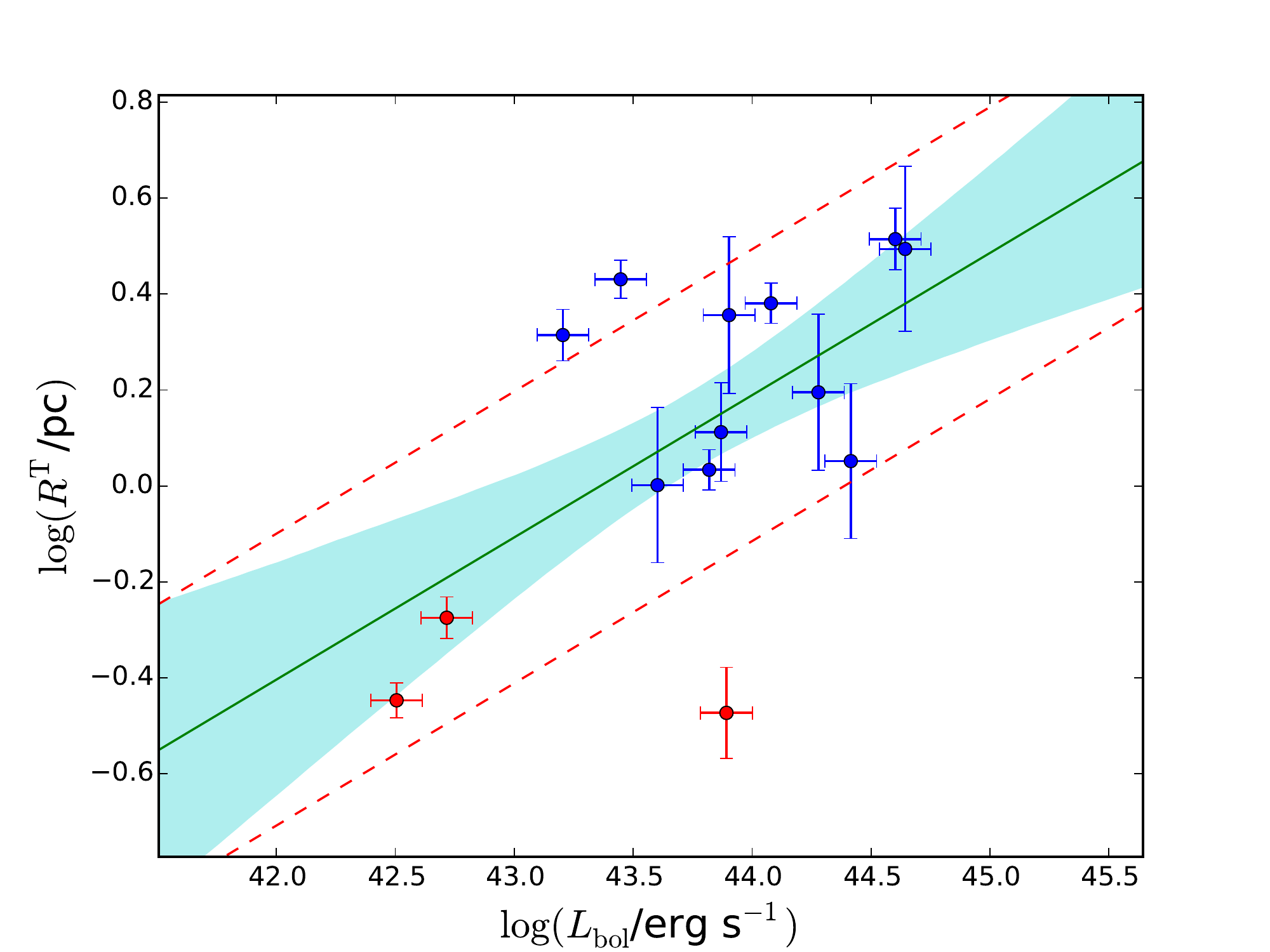}
\label{fig:rout_T_lbol}
   }

\label{fig:subfigurelsize}
\caption{Size-luminosity relations, with progression from an absence of correlation (\ref{fig:rin_lbol}) to a possible one (\ref{fig:rout_T_lbol}). \textit{(a) -} Maser disk inner radius as a function of bolometric luminosity, $L_{\rm bol}$. The three Compton-thin sources are marked with red points. The green line is the best fit linear model, the red dashed lines mark the intrinsic scatter of the data, while the cyan shaded area indicates the uncertainty of the model. \textit{(b) -} Maser disk outer radius as a function of bolometric luminosity, $L_{\rm bol}$. Colors and symbols are the same as in the upper panel. \textit{(c) -} Predicted torus outer radius as a function of bolometric luminosity, $L_{\rm bol}$. Colors and symbols are the same as in the upper panels.}
\end{figure}

\citet{2013A&A...558A.149B} found a clear positive trend of the half-light radius (used as a proxy for the torus size) with the bolometric luminosity, although with large scatter (their Figure 36). In the near-infrared (NIR), a scaling of the dust sublimation radius with $L_{\rm bol}^{1/2}$ is well known. However, this relation is much more scattered in the MIR: more luminous sources generally have larger tori, with no clear trend. We can then explore whether a relation between the bolometric luminosity and the torus size holds in our toy model. We take $L_{\rm bol}$ = $\kappa_{\rm bol}$ $\times$ $L_{2-10}^{\rm int}$, where $\kappa_{\rm bol}$ = 20 $\pm$ 5 is the bolometric correction, constant in our range of intrinsic 2-10 keV luminosities \citep{2006AJ....131.2826S, 2012MNRAS.425..623L}. We choose a 25\% uncertainty on $\kappa_{\rm bol}$ to include also a reasonable error on the 2-10 keV intrinsic luminosity derived from \textit{NuSTAR} spectral fitting, and we note that using a non constant bolometric correction would steepen the correlations. We fitted our sample with a linear relation of the form $\log y = a(\log L_{\rm bol} - 43.5) + b \pm S$, where $L_{bol}$ is measured in erg s$^{-1}$ and $S$ represents the intrinsic scatter in the relation. We applied a Bayesian analysis with loose priors (uniform for all the unknown parameters, i.e. the slope, the intercept, and the intrinsic scatter). 
Since our tori are an extension of the maser disks, we first explored the possibility of a correlation between the inner and outer maser radii with bolometric luminosity. Results are shown in Figure \ref{fig:rin_lbol} and \ref{fig:rout_lbol}. Finally, we repeated the same procedure for the torus outer radius, which is a derived quantity (Figure \ref{fig:rout_T_lbol}); refer to Table \ref{table:lsize} for best fit parameters in all three cases. We find an interesting evolution of the trend, going from the absence of a correlation between the inner maser radius and the bolometric luminosity, to a positive correlation between the torus outer radius and luminosity, although with large intrinsic scatter. The slopes are however all consistent within the uncertainties. \newline The weak trend of the torus outer radius with luminosity could also reflect the weak correlation between maser disks dimensions with luminosity already noted by \citet{2003ApJ...590..162G} comparing NGC 4258 and Circinus inner maser radii. We have here confirmed that finding with an enlarged sample. As already suggested by \citet{2003ApJ...590..162G}, warps in maser disks could break the edge-on geometry condition and bias the disk radial extent measurements.
\begin{table}
\caption{$R_{\rm in}$ - $L_{\rm bol}$, $R_{\rm out}$ - $L_{\rm bol}$ and $R_{\rm out}^{\rm T}$ - $L_{\rm bol}$ relations: best fit parameters.}
\label{table:lsize}
\centering                 
\begin{tabular}{l c c c} 
\hline\hline     
\noalign{\vskip 0.5mm}  
Parameter & $R_{\rm in}$ - $L_{\rm bol}$ & $R_{\rm out}$ - $L_{\rm bol}$ & $R_{\rm out}^{\rm T}$ - $L_{\rm bol}$  \\    
\noalign{\vskip 1mm}   
\hline                        
\noalign{\vskip 1mm}  
$a$ & 0.02$^{+ 0.16}_{- 0.17}$ & 0.12$^{+ 0.14}_{- 0.14}$ & 0.30$^{+ 0.13}_{- 0.13}$ \\ \noalign{\vskip 0.5mm}
$b$ (at $\log L_{\rm bol}/ \text{erg s$^{-1}$} = 43.5$) & - 0.75$^{+ 0.12}_{- 0.12}$ & - 0.29$^{+ 0.09}_{- 0.10}$ & 0.04$^{+ 0.09}_{- 0.09}$ \\ \noalign{\vskip 1mm}  
$S$ [dex] & 0.42$^{+ 0.09}_{- 0.09}$ & 0.34$^{+ 0.07}_{- 0.07}$ &  0.30$^{+ 0.07}_{- 0.07}$ \\ 
\noalign{\vskip 1mm}  
\hline                                 
\end{tabular}
\tablefoot{Errors quoted are 1$\sigma$ confidence level.}
\end{table}
\subsection{The possible role of warps}
\label{sub:warp}
\begin{figure}
\centering
\includegraphics[width=0.5\textwidth]{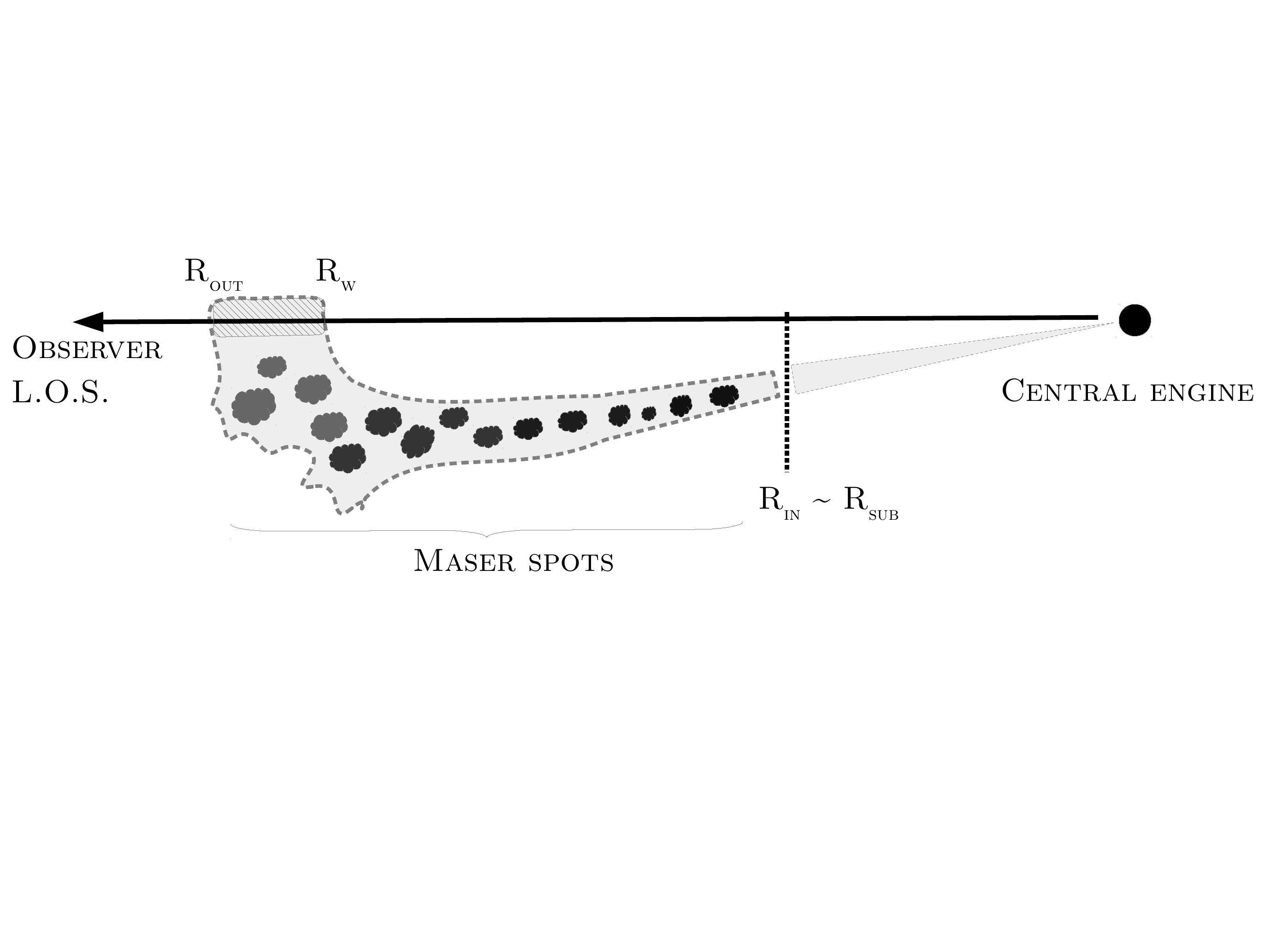}
\caption{Possible role of warps: rapid $N_{\rm H}$ variability and low covering factor. The obscuring column is provided by a warp of depth $\Delta R_{\rm w} = R_{\rm out} - R_{\rm w}$. The thick arrow denotes the observer line of sight.}
\label{fig:sketch2}
\end{figure}

Warped disks have been observed in some sources of the sample (Circinus and NGC 4258 are the clearest cases, see \cite{2003ApJ...590..162G, 2013MNRAS.436.1278W}, and references therein). Low covering factor and/or fast $N_{\rm H}$ variability (like in the case of NGC 4945, see e.g. \cite{2000ApJ...535L..87M, 2014ApJ...793...26P}) could indicate the maser disk as the obscuring structure, instead of invoking an inflated torus. It is easy to see that, keeping the densities expected by the astrophysical maser theory, only a small fraction of the disk is required to intercept the line of sight to have the measured column density (see Figure \ref{fig:sketch2}). To calculate the radial extent of such a warp, we define $R_{\rm w}$ as the warping radius and assume that the warp extends up to the maser outer radius, $R_{\rm out}$. To calculate $R_{\rm w}$, it is sufficient to replace $\rho_{\rm out}$ with $\rho_{\rm w}$ and $\rho_{\rm out}^{\rm T}$ with $\rho_{\rm out}$ in \eqref{eq:torus_rout_exp} and \eqref{eq:nhexp}, if adopting an exponential density profile. Using a power law density profile, results are the same within the uncertainties. \newline In a picture in which there is no standard torus, but a nearly edge-on molecular disk only, a warp of depth $\Delta R_{\rm w} = R_{\rm out} - R_{\rm w}$ is required to obscure the central engine. Numerical values can be found in Table \ref{table:warp}.
\begin{table}
\caption{$\Delta R_{\rm w}$ values needed to have $N_{\rm H, meas}$ = $N_{\rm H, pred}$ with the assumption of all obscuration due to a warp of the maser disk entering the line of sight.}           
\label{table:warp}    
\centering                       
\begin{tabular}{l r r c} 
\hline\hline
\noalign{\vskip 0.5mm}              
Name & $R_{\rm w}$ [pc] & $\Delta R_{\rm w}$ [pc] \\ 
\noalign{\vskip 0.5mm}    
\hline
\noalign{\vskip 1mm}                          
NGC 1194 & 1.29 & 0.04  \\ \noalign{\vskip 0.5mm}  
NGC 1386 & 0.84 & 0.09 \\ \noalign{\vskip 0.5mm}  
NGC 2273 & < 0.13 & > 0.07 \\ \noalign{\vskip 0.5mm}  
NGC 2960 & 0.36 & 0.01 \\ \noalign{\vskip 0.5mm}
NGC 3079 & 1.23 & 0.07 \\ \noalign{\vskip 0.5mm}
NGC 3393 & 1.44 & 0.06 \\ \noalign{\vskip 0.5mm}
NGC 4388 & 0.283 & 0.007 \\ \noalign{\vskip 0.5mm}
IC 2560 & < 0.25 & > 0.08 \\ \noalign{\vskip 0.5mm}
NGC 1068 & < 1.0 & > 0.1 \\ \noalign{\vskip 0.5mm}
NGC 4945 & 0.35 & 0.06 \\ \noalign{\vskip 0.5mm}
Circinus & 0.3 & 0.1 \\ \noalign{\vskip 0.5mm}
NGC 4258 & 0.277 & 0.003 \\ \noalign{\vskip 0.5mm}
NGC 6264 & < 0.77 & > 0.03 \\ \noalign{\vskip 0.5mm}
UGC 3789 & < 0.28 & > 0.02 \\ 
\noalign{\vskip 1mm}
\hline               
\end{tabular}
\tablefoot{Uncertainties are neglected for clarity. Warp depths are lower than, or comparable with, the uncertainties on outer maser radii. The five sources with a lower limit on the column density consequently have a lower limit on the warp depth.}
\end{table}
\section{Conclusions}
\label{sec:conclusions}
We presented hard X-ray spectral analyses of \textit{NuSTAR} data for eight sources out of a sample of 14 nearby disk megamaser galaxies, with the aim of exploring the relationship between the maser disk and the environment in which it resides. In our final sample of 14 AGN, 79\% are Compton-thick, and 21\% are Compton-thin.  All these objects are indeed obscured Sy2 galaxies, and show 22 GHz maser emission from water vapor molecules in a dense molecular disk around active SMBHs.  We proposed a toy model to explain this connection, in which the maser disk is the inner part of the torus, ending in an inflated, geometrically thicker structure. Even if the model is simplistic, is able to recover the column density distribution for a sample of obscured, disk megamaser AGN, using reasonable density profiles (a power law, or, better, an exponential with 1 $\leq$ $n$ $\leq$ 2). Alternatively, one can start from the measured $R_{\rm in}$, $R_{\rm out}$, and $N_{\rm H}$, assume a reasonable density profile, solve the equations for the crucial parameter $\rho_{\rm out}$ and estimate the torus outer radius, which is found to be on the parsec scale. A direct and robust measure of the torus size is available in two sources (NGC 1068 and Circinus) through mid-IR interferometry. In both cases, the outcomes of the model are in agreement with the half-light radius or single resolved structure size measurements. Clearly, a more physical picture explicitly addressing the known disk/torus clumpiness and warping must rely on numerical calculations. \newline 
Assuming a geometry like the one proposed in Figure \ref{fig:torus}, the column density derived with X-ray spectroscopy is due to the inflated end of the clumpy torus only. Indeed, in ten sources of our sample $N_{\rm H} < 10^{25}$ cm$^{-2}$. This obscuration can be explained if the line of sight does not intercept the masing disk. In the remaining five sources, the column density is compatible with that of the maser disk which is, in this framework, very optically thick: these sources could be those seen exactly edge-on, with the line of sight intercepting the geometrically thin maser disk along the equatorial plane. We note, however, that NGC 6264 and UGC 3789 are currently lacking \textit{NuSTAR} spectra and their column density has been estimated using just \textit{XMM-Newton} \citep{2013MNRAS.436.3388C}. Future \textit{NuSTAR} observations may shed light on the exact value of the column density of these two sources, further constraining the fraction of heavily ($N_{\rm H} > 10^{25}$ cm$^{-2}$) Compton-thick AGN in our sample. We finally discussed the possibility that, in some cases, warps in the maser disk may play the role of the classical torus in the AGN unified model, providing low covering factor and fast column density variability.
%
%
\begin{acknowledgements}
We thank the anonymous referee for useful suggestions which helped to improve the paper. This work was supported under NASA Contract NNG08FD60C and made use of data from the \textit{NuSTAR} mission, a  project  led  by  the  California  Institute  of  Technology, managed by the Jet Propulsion Laboratory, and funded by the National Aeronautics and Space Administration. We thank the \textit{NuSTAR} Operations, Software, and Calibration teams for support  with  the  execution  and  analysis  of  these  observations. This  research  has  made  use  of  the
\textit{NuSTAR} Data  Analysis Software  (NuSTARDAS)  jointly  developed  by  the  ASI  Science  Data  Center  (ASDC,  Italy)  and  the  California  Institute of Technology (USA). A.\,M., A.\,C, and S.\,P. acknowledge support  from the ASI/INAF  grant  I/037/12/0-011/13. M.\,B. acknowledges support from NASA Headquarters under the NASA Earth and Space Science Fellowship Program, grant NNX14AQ07H. We acknowledge support from CONICYT-Chile grants Basal-CATA PFB-06/2007 (F.\,E.\,B.\,, C.\,R.\,), FONDECYT 1141218 (F.\,E.\,B.\,, C.\,R.\,), "EMBIGGEN" Anillo ACT1101 (F.\,E.\,B.\,, C.\,R.\,), and the Ministry of Economy, Development, and Tourism's Millennium Science Initiative through grant IC120009, awarded to The Millennium Institute of Astrophysics, MAS (F.\,E.\,B.\,). W.\,N.\,B.\, acknowledges support from NuSTAR subcontract 44A-1092750.
\end{acknowledgements}
\bibliographystyle{aa} 
\bibliography{bib} 
\end{document}